\documentclass[%
 reprint,
 amsmath,amssymb,
 aps,
prd,
]{revtex4-2}

\usepackage{amsmath}
\usepackage{graphicx}% Include figure files
\usepackage{dcolumn}% Align table columns on decimal point
\usepackage{bm}% bold math
\usepackage{hyperref}% add hypertext capabilities
\usepackage{cleveref}
\usepackage{color}
\usepackage{comment}

\usepackage{lipsum}
\usepackage{subcaption}
\usepackage{float}
\usepackage{ulem}
\usepackage{leftidx}
\usepackage{orcidlink}

\newcommand{\dc}[1]{\textsf{\color{magenta}{ #1}}}

\newcommand{\nb}[1]{\textsf{\color{red}{ #1}}}

\begin{document}

\preprint{APS/123-QED}

%\title{Finite temperature effects on fluid oscillations in Neutron Stars}%
\title{Investigating the role of nuclear parameters on oscillation modes\\ in hot Neutron Stars}

\author{Nilaksha Barman\,\orcidlink{0009-0008-1220-359X}}
\email{nilaksha.barman@iucaa.in}
\affiliation{Inter-University Centre for Astronomy and Astrophysics, Pune University Campus, Pune 411007, India}%

\author{Bikram Keshari Pradhan\,\orcidlink{0000-0002-2526-1421}}
\email{bikramp@iucaa.in}
\affiliation{Inter-University Centre for Astronomy and Astrophysics, Pune University Campus, Pune 411007, India}%

\author{Debarati Chatterjee\,\orcidlink{0000-0002-0995-2329}}
\email{debarati@iucaa.in}
\affiliation{Inter-University Centre for Astronomy and Astrophysics, Pune University Campus, Pune 411007, India}%


%\date{}
\begin{abstract}
 Recent studies have revealed that certain nuclear parameters are more dominant than others in governing global neutron star properties, such as its structure or oscillation mode characteristics. Although neutron stars can in general assumed to be cold, in astrophysical scenarios such as newly born neutron stars or remnants of binary neutron star mergers, finite temperature effects play a non-negligible role. In this work, we perform a consistent and systematic investigation of the role of nuclear parameters and thermal effects on neutron star properties and fluid oscillation modes within a full general relativistic scheme. We impose constraints on the parameter space of the relativistic mean field model using state-of-the-art information from terrestrial experiments and multi-messenger astrophysical data. We find effective nucleon mass to be the most important nuclear parameter controlling astrophysical observables of hot neutron stars, similar to the cold beta equilibrated matter. However, we conclude that the interplay among saturation properties and astrophysical observables depends not only on the thermal configurations considered but also on the constraints imposed. We also investigated the role of nuclear saturation parameters on some universal relations for hot NSs which are important in gravitational wave asteroseismology. Our investigation confirmed that these relations are mostly insensitive to nuclear saturation properties and mainly affected by variation of charge fraction in the star. 
\end{abstract}

\keywords{}%Use showkeys class option if keyword
\maketitle

\section{Introduction}\label{sec:intro}

%\dc{What are neutron stars? Why are we interested to study them: cold dense mater, exotic matter}

Neutron stars (NSs) are the densest known compact objects in the observable universe. They provide us unique opportunity to study matter under extreme conditions and hence sometimes referred to as cosmic laboratories. They are formed at the end stage of stellar evolution of massive stars known as core-collapse supernovae or type-II supernovae. Hot newly born neutron stars undergo cooling via neutrino emission and mature into cold degenerate compact objects. Cold NSs are primarily composed of neutrons, protons, electrons and muons  (free neutrinos escape at low temperature $<$ 1 MeV). However, in the high density environment of the NS core other exotic particle degrees of freedom such as hyperons, quarks, kaon condensates etc. might appear in addition to nucleons via weak interactions and  can be in equilibrium with the mixture \cite{Glendenning,Schaffner-Bielich_2020,Lattimer_Prakash_2004science,Lattimer_2012annurev-nucl,Chatterjee_Issac_epja2016}. 
\\

%\dc{How can dense matter in neutron star be modelled? Mention CEFT, RMF}

The composition of the cold NS interior can be described by Equation of State (EoS) models that relate the pressure and the density~\cite{Oertel_Hempel_2017revmodphys}. The EoS connects the microscopic properties of the NS to its global macroscopic structure, such as its mass or its radius. However, the ultradense nuclear matter in the NS core requires solving the complicated many-body problem. There are different approaches e.g. \textit{microscopic} models~\cite{Taranto_Baldo_2013prc,Tong_Wang_2022apj,Schulze_2010_ptps,Drischler_Holt_2010_annurev-nucl} which calculate the nucleon-nucleon interactions ab-initio, while \textit{phenomenological} models~\cite{Hornick_Tolos_2018prc,Typel_Wolter_1999npa} adjust the nucleonic parameters to satisfy observed nuclear saturation properties.
Chiral effective field theory ($\chi$EFT) \cite{Drischler_Carbone_2016prc} is a microscopic method which uses chiral expansion for a reliable description of the sub-saturation region of pure neutron matter. $\chi$EFT results are being used widely in phenomenological or parametric models as a robust method to constrain the parameter space of dense matter EoS at low densities e.g. \cite{Ghosh_Chatterjee_2022epja,Ghosh_Pradhan_2022_fspas,Shirke_Ghosh_2023apj,Beznogov_Raduta_2023prc}.
\\

%\dc{Neutron stars as GW sources: mergers and isolated.}

Being highly relativistic and compact, neutron stars can emit gravitational waves (GWs) via both binary neutron star (BNS) mergers as well as in isolated systems. Any non-axisymmetric perturbations, e.g. quasi-normal modes, deformations such as mountains or magnetic deformations, can lead to time varying mass quadrupole moment in isolated neutron stars and may manifest themselves as gravitational waves. The frequencies of some of the oscillation modes lie in the high frequency band of GWs (e.g. $f$-mode frequencies lie in 1-2 KHz) and provide interesting prospects of detection by current and future generation GW detectors.  The first detection of BNS (GW170817) via GW and electromagnetic (EM) signals opened up a new window in the multimessenger astronomy \cite{LSC_2017_GW170817,LSC_2018_GW170817,LSC_2019_GW170817}. In this context previous works have shown that spin and eccentricity effects lead to excitation of $f$-modes during the inspiral phase of BNS mergers \cite{Chirenti_Gold_2017apj,Steinhoff_Hinderer_2021prr}. With LIGO O4 run and subsequent future GW observations many more BNS merger events are expected to be detected. 
\\

%\dc{How are these models constrained with NS astronomical data}

Any realistic model of neutron stars should satisfy astrophysical constraints from EM and GW observations. The maximum predicted theoretical NS mass should be able to explain the highest observed NS mass of $2.08^{+0.07}_{-0.07}$M$_{\odot}$~\cite{Miller_2021} or $2.078^{+0.067}_{-0.066}$ M$_{\odot}$~\cite{Riley_2021}. EM observations from the recently launched NICER put constraints on masses and radii two pulsars namely PSR J0030-0451 \cite{Miller_2019,Riley_2019} and PSR J0740-6620~\cite{Miller_2021,Riley_2021}. Analysis of GW170817 revealed that the radius of a 1.4M$_{\odot}$ neutron star is $13.4\pm0.2$ km \cite{Tong_Zhao_2020prc} and the estimates for effective tidal deformability for the binary system is $300^{+420}_{-230}$~\cite{LSC_2019_GW170817}. These multi-messenger observations are being used to constrain the high and intermediate density region of nuclear EoSs \cite{Ghosh_Chatterjee_2022epja,Ghosh_Pradhan_2022_fspas,Shirke_Ghosh_2023apj,Beznogov_Raduta_2023prc,De_Finstad_2018prl}. %\dc{Add other references that used GW170817 constraints on NS EOS, e.g. De et al PRL 2017 1804.08583} \nb{added}
\\

%\dc{In which scenarios do we find finite temp effects? SN, PNS, BNS mergers}

Astrophysical events such as proto-neutron stars (PNSs)~\cite{Pons_Reddy_1999apj,Lentz_Bruenn_2015apjl,Janka_Langanke_2007physrep,Burrows_Vartanyan_2020nature} born out of supernovae (SN) and binary neutron star (BNS) mergers~\cite{Baiotti_Rezzolla_2017rpp,Most_Motornenko_2023prd,Fields_Prakash_2023apjl,Radice_Bernuzzi_2020annurev-nucl,Sarin_Lasky_2021gr_grav,Baiotti_2019ppnp} exceed temperatures of tens of MeV and the cold matter approximations are no longer valid.
%    \dc{(cite: Pons+ 1999, Janka+ 2007, Mezzacappa 2015, Burrows+ 2020, Baiotti Rezzolla 2017, Rosswog 2015)}.
It is important to take into account thermal effects as they play a significant role in supernova dynamics, oscillations in PNS as well as in determining stability of the post-merger remnants, which along with the composition  can change the threshold mass limit for prompt collapse to a black hole~\cite{Nunna_Banik_2020_apj}. The EoS of a finite temperature system can be uniquely characterized by three quantities, temperature $T$, baryon number density $n_b$ and the charge fraction of strongly interacting particles $Y_Q$.  In order to be applicable in numerical simulations, a wide range and finely spaced sets of parameters should be considered, along with different possible composition in the neutron star interior. The EoS of hot homogeneous matter spans 0-100 MeV in temperature, 0.01-0.60 in charge fraction and 0.5-6.0 times in nuclear saturation densities in BNS mergers and supernovae~\cite{Oertel_Hempel_2017revmodphys}. In the absence of EoS models spanning the required parameter ranges and composition, the so-called $\Gamma$-law (supplementing the cold EoS with ideal gas thermal contributions) has often been employed to incorporate finite temperature effects in simulations. However, this method of constructing EoS has been shown to be inappropriate particularly for matter consisting of non-nucleonic degrees of freedom \cite{Raduta_2022epja,Kochankovski_Ramos_2022mnras, Blacker_Kochankovski_2024prd,Blacker_Bauswein_2023prd}. It has been described in a recent work in quark-hadron phase transition \cite{Blacker_Bauswein_2023prd} that $\Gamma$ value should be different in hadronic ($\Gamma\sim 1.75$) and quark ($\Gamma\sim 4/3$) phases. The $\Gamma$-law also does not take into account the effect of charge fraction across all densities. Therefore it is crucial to incorporate composition and thermal EoS effects consistently into simulations. \\

%\dc{State of the art: Currently existing finite T EoS. What are their problems}
Currently, there exist only a limited number of available finite temperature EoSs which are widely applied in simulations e.g. Lattimer Swesty (LS220)~\cite{Lattimer_Swesty_1991npa} and Shen et al. (STOS)~\cite{Shen_Toki_1998npa}, both for nucleonic matter and based on single nucleus approximation (SNA). The more recent EoS model of Hempel and Schaffner-Bielich (HS) \cite{Hempel_Schaffner_2010apj} is based on nuclear statistical equilibrium (NSE) which incorporates realistic nuclear masses and goes beyond single-nucleus approximation. The high density core can be modelled using the RMF formalism and is used in constructing non-linear (NL) RMF EoSs such as SFHo/SFHx \cite{Steiner_Hempel_2013apj}, HS(IUF) \cite{Fattoyev_Horowitz_2010prc}, QMC-RMF~\cite{Alford2023}. The RMF EoSs based on linear density dependent (DD) models use the formalism of Typel and Wolter \cite{Typel_Wolter_1999npa}. They include nucleonic EoSs such as HS(DD2) \cite{Hempel_Schaffner_2010apj} as well as models with exotic matter such as hyperons and $\Delta$-baryons e.g. BHB(DD2L$\phi$) \cite{Banik_Hempel_2014apj_supplement}, OMHN(DD2Y) \cite{Marques_Oertel_2017prc}, R(DD2Y$\Delta$) \cite{Raduta_2022epja}. Other finite temperature EoSs include hybrid Chiral Mean field (CMF) model \cite{Papazoglou_Zschiesche_1999prc,Dexheimer_Schramm_2008apj}. 
\\

The existing finite temperature EoS models, which consider a fixed parametrization, correspond to certain values of nuclear saturation properties but they do not span the entire range of uncertainty allowed by current terrestrial experiments. However recent studies have shown that nuclear saturation properties are correlated with astrophysical observables such as radius and tidal deformability in cold neutron stars~\cite{Ghosh_Chatterjee_2022epja,Ghosh_Pradhan_2022_fspas,Pradhan_Chatterjee_2022prc,Pradhan_Chatterjee_2023npa,Thapa_Beznogov_2023prd,Beznogov_Raduta_2023prc}. These studies provide an interesting prospect to investigate the role of nuclear saturation effects on NS observable properties at finite temperature. There have been a few recent works in this direction. An EoS formalism within the covariant density functional approach with density dependent couplings employing the posteriors of a Bayesian inference (DDB*) formalism~\cite{Beznogov_Raduta_2023prc} has been extended to finite temperature by Raduta et al.~\cite{Raduta_Beznogov_2024_plb}. In another approach, the effect of slope of symmetry energy $L_{sym}$ and skewness $Q_{sat}$ on finite temperature EoS has been studied within the covariant density functional scheme for fixed DD-RMF parametrizations in ~\cite{Tsiopelas_Sedrakian_2024,Li_Sedrakian_2023apj}. 
\\

Significant efforts have been dedicated for advancing our understanding of GW asteroseismology for PNSs, through the study of GWs emitted by PNS oscillation modes~\cite{Sotani2016,Sotani:2019b,Torres2018}. Both   simulations~\cite{Torres2018,Torres2019,Sotani2020c,Rodriguez2023,Sotani2024a} and perturbative approaches~\cite{Sotani2020,Thapa_Beznogov_2023prd} have been considered in studying the PNS oscillations.
In the simulation by Torres et al.~\cite{Torres2018}, it was shown that most of the GW energy is stored in the lowest-order eigenmodes, in particular $\leftidx{^2}{g_1}$ and $\leftidx{^2}{f}$. The same group in their previous study in Cowling approximation showed that higher order $p_i$-modes upto 5 nodes and $g_1$-modes appear in GW spectrogram \cite{Torres_Durán_2017mnras}. PNS simulations by Sotani et al.~\cite{Sotani2020,Sotani2020b,Sotani2020c} also indicate that $g_1$ and $f$-modes appear to dominate GW spectra in the earlier and later stages of PNS evolution respectively.
 Further analyses involve assessing the detectability of GWs from PNSs and inferring PNS properties from the GW data~\cite{Bruel2023,Bizouard2020}.
 \\

Cold NSs are known to follow certain EoS independent universality such as $C$-Love, $\bar{I}-\Lambda-\bar{Q}$, $f$-Love~\cite{Maselli_Cardoso_2013prd,Yagi_Yunes_2013prd,Pradhan_Chatterjee_2022prc,PradhanMNRAS,PradhanPRD,Wen2019}~among others. Much effort has been dedicated into deriving universal relations (URs) that link the oscillation mode properties of PNSs with their physical characteristics, such as mass and radius, which are crucial for GW asteroseismology~\cite{Torres2019,Sotani:2019b,Sotani2024a,Li:2019xxz} of proto-neutron stars. Raduta et al.~\cite{Raduta_Oertel_2020mnras} obtained $\Bar{I}-\Lambda-\Bar{Q}-C$ relations for hot NSs using a DD-RMF parameterization.  This was extended by Khadkikar et al.~\cite{Khadkikar_Raduta_2021prc} to explore URs between static and Kepler rotation properties such as mass, radius and frequency etc. and   fit relations were provided for $\Bar{I}-C$ and $\Bar{Q}-C$ for different thermal configurations with various tabulated EoSs. Similar UR study for nucleonic and hybrid hot stars were performed by Largani et al.~\cite{Largani_Fischer_2022mnras} where they provided fits for normalized moment of inertia with compactness at Keplerian limit for fixed values of entropy per baryon and lepton fraction. Laskos-Patkos et al.~\cite{Laskos-Patkos_Koliogiannis_2022_universe} studied deviations of hot neutron stars from cold universality for $C$-Love relations.
\\

In a series of recent works~\cite{Ghosh_Chatterjee_2022epja,Ghosh_Pradhan_2022_fspas,Shirke_Ghosh_2023apj}, we developed a formalism to constrain the EoS of cold dense NS matter using $\chi$EFT at low densities, heavy-ion data at intermediate densities and multi-messenger astrophysical data at high densities, within a non-linear relativistic mean field model (NL-RMF) framework. These works also investigated the correlation of nuclear empirical parameters with NS macroscopic properties (radius $R$, tidal deformability $\Lambda$) and found the effective nucleon mass $m^*$ to be strongly correlated to these observables. Further, recent investigations within the same formalism also showed that $m^*$ is the dominant parameter related to oscillation modes such as fundamental ($f$-) mode in NSs~\cite{Jaiswal2021,Pradhan_Chatterjee_2021prc,Pradhan_Chatterjee_2022prc}. However, it is not clear whether such correlations will hold in finite temperature scenarios, such as oscillations in newly born neutron stars or binary neutron star mergers. A few interesting studies recently probed thermal effects on the NS EoS imposing constraints from ${\chi}$EFT and NS maximum mass within the NL-RMF formalism (see e.g.~\cite{Alford2023}). In another recent work~\cite{Thapa_Beznogov_2023prd}, thermal effects were studied by considering a few density-dependent (DD) RMF tabulated EoSs from CompOSE~\cite{CompOSE}. The effect of finite temperature on $f$- and first overtone pressure ($p_1$)-modes was studied within Cowling approximation for the chosen EoSs. However, the correlations among nuclear and NS properties were investigated only for cold NS matter employing the DDB* model. From these studies, the correlations of the nuclear saturation properties on NS global properties and general relativistic mode characteristics at finite temperature are not directly evident. 
\\

%\dc{Aim of this paper}
The aim of this work is to perform a systematic and consistent investigation of the role of nuclear saturation parameters as well as thermal properties i.e., entropy per baryon, $S/A$ and out-of-$\beta$ equilibrium condition (charge/electron fraction, $Y_Q$) on the EoS of hot and dense matter. For this we extend our recently developed formalism of imposing constraints at different densities on the parameter space in the RMF scheme for cold matter EoS to incorporate thermal effects.  The finite temperature EoS is then used to derive macroscopic properties of hot neutron stars.  The EoS framework is further applied for the study of oscillation modes within a fully general relativistic treatment. With the posteriors, we search for possible correlations between nuclear saturation parameters as well as thermal properties with astrophysical observables. We also investigate the effect of nuclear saturation and thermal properties on the Universal relations. Note that we have not included the effects of neutrino trapping in our study. 
%and to compare these results with previous works involving cold matter \cite{Pradhan_Chatterjee_2021prc,Pradhan_Chatterjee_2022prc} and  DD-RMF models (cold and hot cases) \cite{Thapa_Beznogov_2023prd}.
\\

%\dc{Structure of this paper}

The paper is structured as follows: In section \ref{sec:formalism} we provide a description of the formalism involved. In subsection \ref{subsec:EOS} the RMF model and its extension to include thermal effects have been given. We also derive the expressions of various thermodynamic quantities of interest. In the next subsection \ref{subsec:macro} we determine macroscopic NS properties, while in the subsection \ref{subsec:modes} the fully General Relativistic (GR) description of calculating the oscillation modes are given. In section \ref{sec:results} we discuss results for thermal effects on EoSs consistent with $\chi$EFT, astrophysical and heavy ion collision constraints as shown in subsection~\ref{subsec:EoS_res} and their global properties in subsection~\ref{subsec:M-R_res}. 
%The former represent a less compact configuration as we shall see in subsection \ref{subsec:M-R_res}.
We then discuss results for oscillation mode frequencies and damping timescales in \ref{subsec:f-mode_res}. We also present the results of the investigation of correlations of the nuclear properties with observable NS properties.
%We show how the nuclear saturation properties affect the observables in the two cases as correlation matrices. 
In \ref{subsec:UR} we show deviation of universality of oscillation modes at finite temperature compared to cold $\beta$ equilibrated matter. Then we summarise our results in section \ref{sec:discussions}. Throughout the paper we use the natural units ($c=\hbar=k_B=G=1$).

\section{Formalism}
\label{sec:formalism}
%\dc{Microscopic EoS model: finite T NL-RMF, Lagrangian with equations}

\subsection{Equation of State}\label{subsec:EOS}
The EoS of the NS core is modelled using a non-linear relativistic mean field (NL-RMF) formalism ~\cite{Chen_Piekarewicz_2014prc,Hornick_Tolos_2018prc}. The nucleon interactions occur via exchange of $\sigma$, $\omega$ and $\Vec{\rho}$ mesons. $\sigma$ is a scalar meson that facilitates long range attraction while $\omega$ is a vector meson for nucleon-nucleon repulsion. $\Vec{\rho}$ vector meson is responsible for isospin asymmetry and to maintain charge neutrality. The non-linear terms are the scalar self-interaction term $U_{\sigma}$  given by $U_{\sigma} = \frac{1}{3}b m_N (g_{\sigma N} \sigma)^3 + \frac{1}{4}c (g_{\sigma N} \sigma)^4$, and $\Lambda_{\omega}$ which mimics the $\omega-\rho$ interaction.
The Lagrangian density for the homogenous nuclear matter is given below: 
\begin{align}\label{eq:NL_RMF}
    \mathcal{L} &= \sum_{i \in B}\Bar{\psi}_i (i \gamma_{\mu} \partial^{\mu} - m_i + g_{\sigma i} \sigma - g_{\omega i} \gamma^{\mu} \omega_{\mu} - g_{\rho i} \gamma^{\mu} \rho_{\mu}^a \tau^a )  \psi_i \nonumber \\
    &+ \frac{1}{2} \partial^{\mu} \sigma \partial_{\mu} \sigma - \frac{1}{2} m_{\sigma}^2 \sigma^2 - \frac{1}{4} \omega_{\mu \nu} \omega^{\mu \nu} + \frac{1}{2} m_{\omega}^2 \omega^2 \nonumber \\
    &- \frac{1}{2} \Vec{\rho}_{\mu \nu} \Vec{\rho}^{\mu \nu} + \frac{1}{2} m_{\rho}^2 \Vec{\rho}^2 + \Lambda_{\omega} (g_{\rho N}^2 \Vec{\rho}_{\mu}.\Vec{\rho}^{\mu}) (g_{\omega N}^2 \omega_{\mu} \omega^{\mu}) \nonumber \\
    &- U_{\sigma} + \Bar{\psi}_e (i \gamma_{\mu} \partial^{\mu} - m_e) \psi_e \nonumber \\
    &+\sum_{i = B,e}\Bar{\psi}_i(-q_i \gamma_{\mu} A^{\mu})\psi_i -\frac{1}{4}F_{\mu\nu}F^{\mu\nu}~,
\end{align}
where, $\psi_i$ are the Dirac spinor for nucleon(s). The isoscalar couplings ($g_{\sigma B},g_{\omega B}, b, c$) are fitted to reproduce isoscalar saturation properties (saturation density $n_{sat}$, energy per particle $E_{sat}$, compressibility $K_{sat}$, effective nucleon mass $m^*$) while isovector coupling constants ($g_{\rho B},~\Lambda_{\omega}$) 
 are fitted to reproduce isovector saturation properties (symmetry energy $J_{sym}$ and its slope $L_{sym}$). The leptonic species present are non-interacting electrons ($e^-$) obeying Fermi-Dirac (FD) statistics; muons ($\mu^-$) are not considered as they are expected to appear after the star reaches $\beta$ equilibrium. Photons ($\gamma$) are bosonic non-interacting degrees of freedom described by $A_{\mu}$ and $F^{\mu \nu} = \partial^{\mu} A^{\nu} - \partial^{\nu} A^{\mu}$. Under RMF approximation the meson mean field equations become,
\begin{align}
    m_{\sigma}^2 \bar{\sigma} &= \sum_{i \in B} g_{\sigma i} n_{i}^s - \frac{\partial U}{\partial \bar{\sigma}} \label{eq:sigma} \\
    m_{\omega}^2 \bar{\omega}_0 &= \sum_{i \in B} g_{\omega i} n_{i} - 2 \Lambda_{\omega} g_{\rho}^2 g_{\omega}^2 \bar{\rho}_{03}^2 \bar{\omega}_0   \label{eq:omega} \\
    m_{\rho}^2 \bar{\rho}_{03} &= \sum_{i \in B}  g_{\rho i} t_{3i} n_{i} - 2 \Lambda_{\omega} g_{\rho}^2 g_{\omega}^2 \bar{\rho}_{03} \bar{\omega}_0^2 \label{eq:rho}~,
\end{align}
where $\bar{\sigma}$, $\bar{\omega_0}$, $\bar{\rho_{03}}$ are the mean meson fields and $t_{3i}$ is the third component of the isospin of the $i$-th baryonic species.
\\

Including thermal effects, the number density and scalar density of the $i$-th baryonic species respectively are~\cite{Raduta_2022epja,Raduta_Beznogov_2024_plb}:
\begin{align}
    n_i &= \frac{2J_i+1}{2\pi^2} \int dk k^2 \left[f_{FD} \left(\frac{E_i(k) - \mu_i^*}{T}\right)\right. \nonumber \\
    &\left. - f_{FD} \left(\frac{E_i(k) + \mu_i^*}{T}\right) \right] \label{eq:number_den}~, \\
    n_i^s &= \frac{2 J_i+1}{2\pi^2} \int dk \frac{k^2 m_i^*}{E_i(k)} \left[f_{FD} \left(\frac{E_i(k) - \mu_i^*}{T}\right) \right. \nonumber \\
    &\left .+ f_{FD} \left(\frac{E_i(k) + \mu_i^*}{T}\right) \right] \label{eq:scalar_den}~.
\end{align}
In eqs.~\ref{eq:number_den} and \ref{eq:scalar_den}, $m_i^* = m_i - g_{\sigma i}\bar{\sigma}$ is the effective baryon mass and $E_i(k) = \sqrt{k^2+{m_i^*}^2}$, the effective energy dispersion of the $i$-th baryonic species. $J_i$ is the spin of the $i$-th baryonic species. $f_{FD}(x)=1/(1+e^x)$ is the FD distribution function, $T$ is the equilibrium temperature of the mixture and $\mu_i^*$ is the effective chemical potential of the $i$-th species given by,
\begin{equation}\label{eq:eff_chem_pot}
    \mu_i^* = \mu_i - g_{\omega i}\bar{\omega}_0 - g_{\rho i} t_{3i} \bar{\rho}_{03}~.
\end{equation}
where $\mu_i$ is the actual chemical potential of the corresponding species and can be written in terms of baryon ($\mu_B$) and charge ($\mu_Q$) chemical potentials. If $B_i$ and $Q_i$ are baryon and electric charges of the $i$-th species then, $\mu_i = B_i \mu_B + Q_i \mu_Q$. Both particle and anti-particle species contributes to entropy, as do leptons/anti-leptons and radiation. The effective chemical potentials for anti-particles are negative of corresponding particle effective chemical potentials. Chemical potential for electrons is $\mu_e = \mu_L-\mu_Q$, where $\mu_L$ is the lepton chemical potential. The total entropy density ($s$) is given by,
\begin{align}\label{eq:entropy_den}
    s &= -\sum_{i \in B} \frac{2 J_i+1}{2\pi^2} \int dk k^2 \left[f_{FD}\left(\frac{E_i(k) - \mu_i^*}{T}\right)\right. \nonumber \\
    &\left .\ln f_{FD}\left(\frac{E_i(k) - \mu_i^*}{T}\right) + \Bar{f}_{FD}\left(\frac{E_i(k) - \mu_i^*}{T}\right)\right. \nonumber \\
    &\left . \ln \Bar{f}_{FD}\left(\frac{E_i(k) - \mu_i^*}{T}\right)\right] \nonumber \\
    &+ anti-baryon~contribution~(\mu^*_i\rightarrow-\mu^*_i) \nonumber \\
    &+ s_{e^-} + s_{e^+} \nonumber \\
    &+ s_{\gamma}~,
\end{align}
%\bkp{In ~\cref{eq:entropy_den} what is $s_{\gamma}$, I do not see any electromagnetic contribution in the Lagrangian~\eqref{eq:NL_RMF}? Sorry if I missed anything.} \nb{done} \dc{What are the $s$ and where are they defined?} \nb{done}
where, $\Bar{f}_{FD}(x) = 1 - f_{FD}(x)$ is the Fermi-Dirac distribution function for vacancy of state.
The electronic contribution to the entropy density ($s_{e^-}$) is found by substituting electron chemical potential ($\mu_e$) in place of effective chemical potential ($\mu_{i}^*$) in eq.~\ref{eq:entropy_den}, while the radiation ($\gamma$) contribution is $s_{\gamma} = 4 \pi^2 T^3 / 45$. The entropy per baryon is then given by $S/A=s/n_B$. Also radiation energy density, $\epsilon_{\gamma} = \pi^2 T^4 / 15$. From the Lagrangian described in eq.~\ref{eq:NL_RMF}, the energy density can be evaluated from the energy momentum tensor $T^{\mu \nu}$ in the mean field approximation, and the pressure can be evaluated using the Gibbs-Duhem relation, as given below:

%\bkp{Before going to the result section, here may be the plots for the EOS, temperature evolution for a given  RMF parametrization with different choices of ($S/A$, $Y_Q$) can be added.} \nb{done} \\

\begin{align}\label{eq:EoS} 
    \epsilon &= <T^{00}> \nonumber \\
    &= \frac{1}{2} m_{\sigma}^2 \Bar{\sigma}^2 + \frac{1}{2} m_{\omega}^2 \Bar{\omega}_0^2 + \frac{1}{2} m_{\rho}^2 \Bar{\rho}_{03}^2 + U_{\sigma}  \nonumber \\
    &+ 3\Lambda_{\omega} (g_{\rho N} g_{\omega N} \Bar{\rho}_{03}.\Bar{\omega}_{0})^2 \nonumber \\
    &+ \sum_{i} \frac{2 J_i+1}{2\pi^2} \int dk k^2 E_{i}(k) \left [f_{FD}\left(\frac{E_i(k) - \mu_i^*}{T}\right)\right. \nonumber \\
    &+\left. f_{FD}\left(\frac{E_i(k) + \mu_i^*}{T}\right) \right] \nonumber \\
    &+ \epsilon_{e^-} + \epsilon_{e^+} \nonumber \\
    &+ \epsilon_{\gamma}~, \\
    p &= \sum_{i = B, e} \mu_i n_i + Ts - \epsilon~.
\end{align}
The electron contribution to the energy density is again found by replacing effective chemical potential ($\mu_i^*$) with electron chemical potential ($\mu_e$). For describing the low density part of the EoS the HS(DD2) model is used~\cite{Hempel_Schaffner_2010apj}, taken from CompOSE database \cite{CompOSE}. For each thermal case, the crust and core are matched by keeping $S/A$ and $Y_Q$ same for crust [HS(DD2)] and core (NL-RMF) ensuring thermodynamic consistency.
\\

The direct consequences of finite temperature are effects such as out-of-$\beta$ equilibrium, neutrino trapping among others. These effects can manifest themselves in the hot EoS and consequently in the macroscopic properties. At postbounce evolution in PNSs, the thermal properties control the EoS until the neutron star becomes cold and degenerate. These thermal properties are entropy per baryon ($S/A$) and electric charge (out-of-$\beta$ equilibrium) and/or lepton fractions (in case of neutrino trapping). During PNS evolution and also BNS merger scenarios, the system goes through different stages of $S/A$ and $Y_Q$. We discuss two such thermal configurations relevant for PNS evolution in section~\ref{sec:results} and show the effect of variation of nuclear properties on these configurations subsequently. It has been shown in previous works that this leads to deviations from universal relations (URs) usually observed in cold neutron stars \cite{Raduta_Oertel_2020mnras,Laskos-Patkos_Pavlos_2023hnpsanp,Marques_Oertel_2017prc}.
\\

\begin{figure}[H]
    \includegraphics[width=\linewidth]{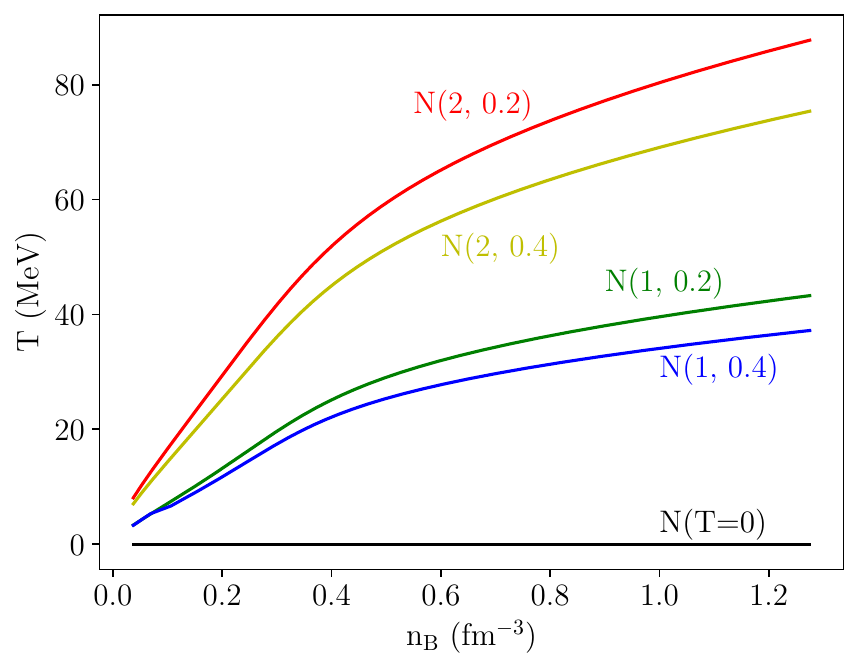}
    \caption{Temperature profiles in the NS core, for thermodynamic conditions $N(S/A = 1,~Y_Q=0.2)$,~$N(S/A = 1,~Y_Q=0.4)$,~$N(S/A = 2,~Y_Q=0.2)$ and $N(S/A = 2,~Y_Q=0.4)$. The saturation properties have been fixed to 
    $n_{sat}=0.15~\rm{fm^{-3}}$, $E_{sat}=-16.0$ MeV, $K_{sat}=240$ MeV, $J_{sym}=32$ MeV, $L_{sym}=60$ MeV, $m^*/m=0.65$.}
    \label{fig:T_vs_nb}
\end{figure}

To verify expected behaviour of thermal properties as given in existing literature, we first use a fixed nuclear parameterization (from table~\ref{tab:prior}) and vary only the thermal properties of the star. The core temperature profiles corresponding to different thermodynamic conditions for nucleonic matter $N(S/A,Y_Q)$ i.e., different values of entropy per baryon $S/A$ and charge fraction $Y_Q$, are shown in fig. \ref{fig:T_vs_nb}. It can be seen that higher entropy leads to larger temperatures and higher proton fraction results in lower temperatures at high densities.

\subsection{Macroscopic and Tidal properties}
\label{subsec:macro}

We assume a perfect fluid configuration which is static and spherically symmetric. The space time geometry for such a configuration is given by Schwarzschild interior metric 
\begin{equation}\label{eq:metric}
    ds^2 = -e^{2\Phi (r)} dt^2 + e^{2\lambda (r)} dr^2 + r^2 (d\theta^2 + sin^2\theta d\phi^2)~.
\end{equation}
consisting of metric functions $\Phi (r)$ and $\lambda (r)$ which depend on the underlying equation of state $[p = p(\epsilon)]$.
The macroscopic configuration of the hot neutron star is determined by the hydrostatic equilibrium condition and is obtained by solving the Tolman-Oppenheimer-Volkoff (TOV) equations:
\begin{align}\label{eq:TOV}
    \frac{dm (r)}{dr} &= 4 \pi r^2 \epsilon(r)~, \\
    \frac{dp (r)}{dr} &= -[p(r)+\epsilon(r)] \frac{[m(r)+4\pi r^3 p(r)]}{\left[r^2\left(1-\frac{2m(r)}{r}\right)\right]}~, \\
    e^{2 \lambda (r)} &= \frac{1}{\left[1-\frac{2m(r)}{r}\right]}~, \\
    \frac{d \phi(r)}{dr} &= - \frac{1}{\left[\epsilon (r) + p(r)\right]} \frac{dp (r)}{dr}~.
\end{align}
Here $m(r)$ is the mass enclosed within a spherical shell of radius $r$ and therefore if $R$ is the radius of the star then $m(r=R) = M$, the NS mass. The tidal properties of the star are encoded in tidal love number $k_2$ \cite{Yagi_Yunes_2013prd} and it is related to dimensionless tidal deformability $\Lambda$ through stellar compactness $C = M/R$.
\begin{equation}\label{eq:tidal_deform}
    \Lambda = \frac{2}{3} \frac{k_2}{C^5}~.
\end{equation}

\begin{figure}[H]
    \centering
    \includegraphics[width=.95\linewidth]{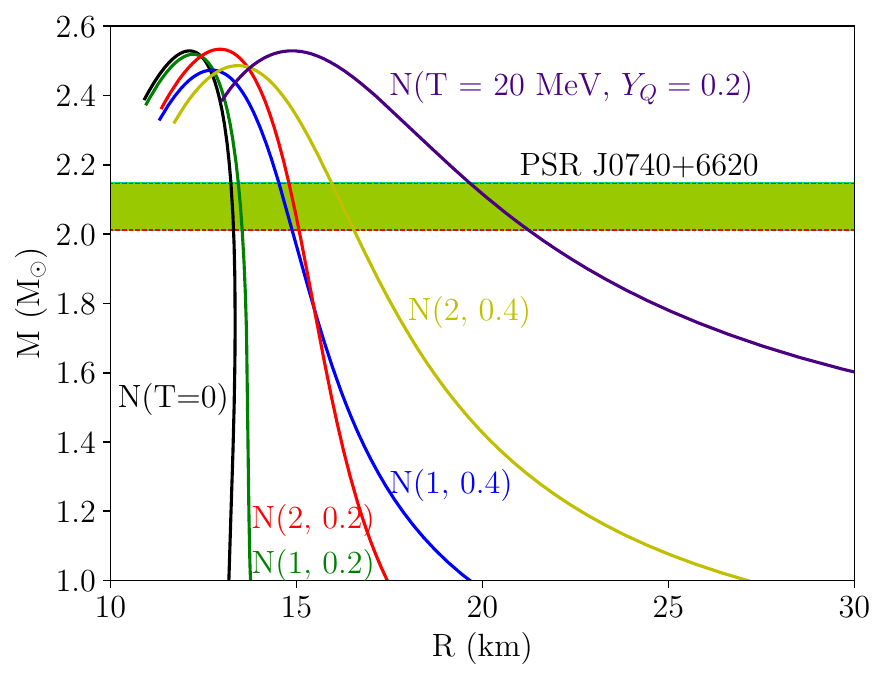}
    \caption{Mass-Radius relations for the different thermal configurations in fig.~\ref{fig:T_vs_nb}.} 
    \label{fig:m_vs_r}
\end{figure}

We display the NS mass-radius relations in fig.~\ref{fig:m_vs_r} corresponding to the thermodynamic conditions mentioned in the previous section. The case $N(T=0)$ stands for $\beta$ equilibrated nucleonic matter.
It can be observed from fig.~\ref{fig:m_vs_r} that the increase of charge fraction leads to a decrease in maximum mass supported by the hot neutron star. Both the increase in $S/A$ and $Y_Q$ results in higher radii for the same mass configurations in intermediate mass ranges. As an example we have also shown the isothermal configuration with $T = 20~\rm{MeV}$ and $Y_Q=0.2$. For this configuration low density part is taken again from HS(DD2) EoS with $T = 20~\rm{MeV}$ and $Y_Q=0.2$ from CompOSE~\cite{CompOSE} to ensure thermodynamic consistency. This configuration has the flattest mass-radius relations among all the cases.

\subsection{Oscillation modes}
\label{subsec:modes}
In general relativity (GR), NS oscillation modes are analyzed by considering perturbations in both the fluid and the background metric. These modes are classified into polar and axial types depending on their transformation under parity~\cite{Thorne}. Our focus here is on polar (which are strongly coupled to the fluid) and non-radial (dominant in the GW emission) oscillations. The theory of non-radial oscillations of non-rotating NSs, along with the mathematical formalism in the context of GR, was developed in the pioneering work of Thorne~\cite{Thorne}. In GR, these modes are quasi-normal, characterized by complex eigenfrequencies ($\omega$). The real part of $\omega$ corresponds to the mode frequency ($\nu$), while the imaginary part is related to the gravitational damping time ($\tau$).\\

A simplified approach known as the relativistic Cowling approximation, which neglects metric perturbations, is frequently used to solve for the oscillation modes, providing information solely on the oscillation frequency. While GR offers a comprehensive understanding of oscillation modes and GW damping times, it is more complex and computationally involved. Meanwhile, the simplified relativistic Cowling approximation has been used to draw many qualitative conclusions and study PNS oscillations. The accuracy of the oscillation mode solutions in the relativistic Cowling approximation of the PNS has been demonstrated in~\cite{Sotani2020}. Similar to the cold NSs, the relativistic Cowling approximation can overestimate the $f$-mode frequency by $\sim$20 \% compared to that obtained in the GR formalism. Further investigations involve searching for the dependence of the mode properties on the outer boundary conditions~\cite{sotani2019} and numerical dimensions~\cite{Sotani2020c}. Additionally,  in the work of~\cite{Torres2018}, the post-bounce oscillation modes of PNSs have also been studied in general relativity, including spacetime perturbations.

Several approaches have been developed in the past few decades to solve for the oscillation modes within the linearized theory of GR, including the direct integration method~\cite{Detweiler83, Detweiler85}, WKB approximation, the CF method by Chandrasekhar and Ferrari~\cite{Chandrasekhar:1991} and the method of continued fraction~\cite{Sotani2001, Leins1993}. The differential equations for the fluid and metric variables for solving the non-radial modes can be found in~\cite{Sotani2020, Pradhan_Chatterjee_2022prc}. We adopt the direct integration method developed by  Lindblom and  Detweiler~\cite{Detweiler85} to solve for the non-radial QNM modes. In short, the coupled perturbation equations for perturbed metric and fluid variables are integrated within the NS interior with appropriate boundary conditions. We define the PNS boundary as the radius obtained from the TOV solutions. Further, the fluid variables are set to zero outside the star, and the Zerilli wave equation~\cite{Zerilli} is integrated far away from the star. Further, a search is carried out for the complex mode frequency ($\omega=2\pi \nu+\frac{1}{\tau}$) corresponding to only the outgoing wave solution to the Zerilli equation at infinity. For finding the mode characteristics, we use the numerical methods developed in our previous work~\cite{Pradhan_Chatterjee_2022prc}.

\begin{figure}[H]
    \centering
    \begin{subfigure}[b]{0.5\textwidth}
        \includegraphics[width=\linewidth]{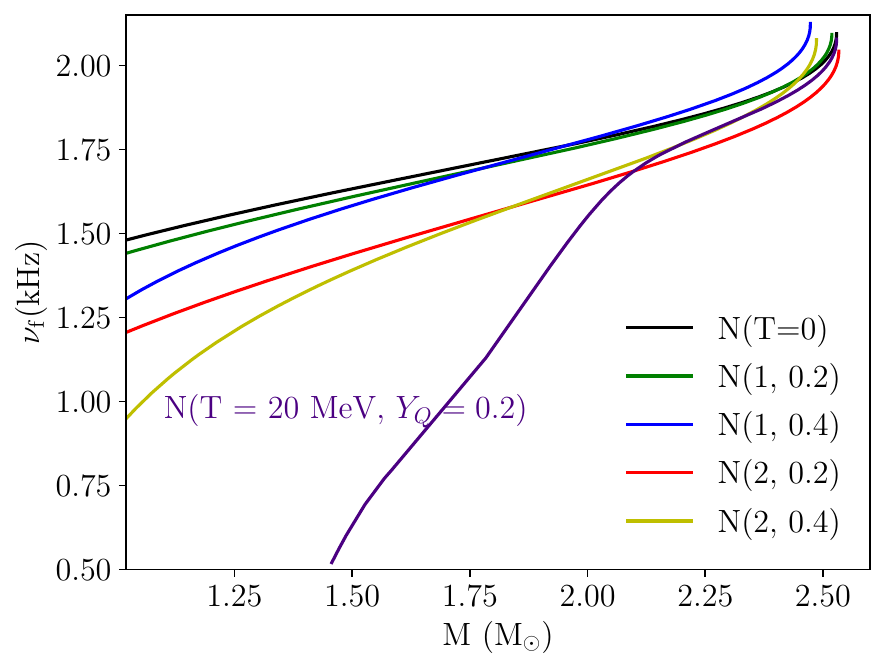}
        \caption{$f$-mode frequency v/s mass}
        \label{subfig:f_vs_m}
    \end{subfigure}
    \hfill
    \begin{subfigure}[b]{0.5\textwidth}
        \includegraphics[width=\linewidth]{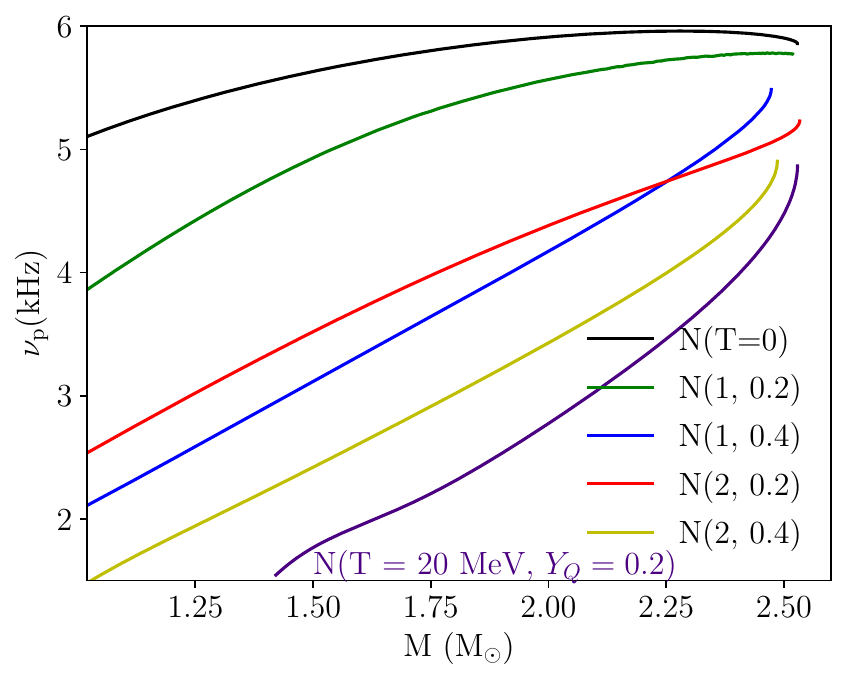}
        \caption{$p_1$-mode frequency v/s mass}
        \label{subfig:p1_vs_m}
    \end{subfigure}
    \caption{$f$($p_1$)-mode frequencies as function(s) of mass. The nuclear saturation properties are the same as in fig.~\ref{fig:T_vs_nb}.}
    \label{fig:f/p1_vs_m}
\end{figure}

We display $f$- and $p_1$-mode frequencies of some specific thermal cases in Fig.~\ref{subfig:f_vs_m} and~\ref{subfig:p1_vs_m} respectively calculated in general relativistic treatment. The $p_1$-mode frequencies are highest for the cold configuration and they are found to decrease with an increase in entropy per baryon and charge fraction. The isothermal configuration for $T=20$ MeV has the lowest $p_1$-mode frequencies. A crossing appears between $N(1, 0.4)$ and $N(2, 0.2)$ configurations for the $p_1$-mode at high mass. This behavior is also seen in the full parameter space allowed by various constraints (discussed in subsection~\ref{subsec:f-mode_res}). On the other hand, $f$-mode frequencies for all cases have some overlapping regions at intermediate and high mass ranges. From ~\cref{fig:f/p1_vs_m}, it can be concluded that both the $f$- and $p_1$-mode frequencies decrease in thermally configured stars compared to cold NSs. Notably, the thermal effects have a more pronounced impact on the $p_1$-mode frequencies than on the $f$-mode frequencies. The reduction in mode frequencies with thermal effects aligns with the findings of previous studies ~\cite{Thapa_Beznogov_2023prd, Burgio_Ferrari_2011prd, Ferrari2003}. This behavior can be attributed to the fact that both $f$-mode and $p$-mode frequencies scale inversely with a power of the stellar radius (see~\cite{Sotani_Kuroda_2017prd,Sotani:2019b}). As thermally configured stars of a given mass exhibit lower compactness than their cold counterparts, the increase in radius leads to a reduction in mode frequencies. Thus, the observed variations in frequency are a direct consequence of the underlying differences in compactness between hot and cold configurations (see~\cref{fig:m_vs_r}).

\begin{comment}

\begin{figure}[H]
    \centering
    \begin{subfigure}[b]{0.5\textwidth}
        \includegraphics[width=\linewidth]{f_vs_m.pdf}
        \caption{$f$-mode frequency v/s mass}
        \label{subfig:f_vs_m}
    \end{subfigure}
    \hfill
    \begin{subfigure}[b]{0.5\textwidth}
        \includegraphics[width=\linewidth]{p1_vs_m.pdf}
        \caption{$p_1$-mode frequency v/s mass}
        \label{subfig:p1_vs_m}
    \end{subfigure}
    \caption{$f$($p_1$)-mode frequencies as function(s) of mass. The nuclear saturation properties are the same as in fig.~\ref{fig:T_vs_nb}.}
    \label{fig:f/p1_vs_m}
\end{figure}

\end{comment}

\begin{comment}
\begin{figure}[H]
    \centering
    \includegraphics[width=\linewidth]{f_vs_m.pdf}
    \caption{f-mode frequencies v/s Mass in Cowling Approximation for the different thermal configurations in fig.~\ref{fig:T_vs_nb}.}
    \label{fig:f_vs_m}
\end{figure}
\end{comment}

\subsection{Constraining the model parameters}
\label{subsec:bayesian}

%\dc{Please rewrite this section in further detail, describing the Bayesian scheme in brief within 1-2 paragraphs. See Suprovo's publications, e.g. Shirke et al. 2023 ApJ 944 7, Sec 2.3.} \nb{This subsection has been updated.}\\
The parameter space of the EoS considered should be consistent with microscopic calculations, state-of-the-art astrophysical observations and terrestrial nuclear experiments. In previous works by some of the authors of this paper~\cite{Ghosh_Chatterjee_2022epja,Ghosh_Pradhan_2022_fspas}, a Bayesian-like cut-off scheme described in this section has been used to restrict the parameter space of EoS of cold nucleonic and hyperonic NS matter by imposing the above constraints. In Shirke et al. (2023)~\cite{Shirke_Ghosh_2023apj}, this formalism was extended to cold hybrid stars with phase transition using constraints from perturbative QCD (p-QCD). In this work, we have applied a similar scheme to extend the study to finite temperature, as described below.
\begin{enumerate}
    \item \textbf{Chiral Effective Field Theory ($\chi$EFT):} The $\chi$EFT takes into account nuclear many body interactions using order-by-order expansion in terms of contact interactions and long-range pion exchange interactions and low momentum expansion of nuclear forces related to QCD symmetries are applied. These are based on microscopic calculations from the works of Drischler~et~al. \cite{Drischler_Carbone_2016prc,Drischler_Hebeler_2019prl}. These constraints can been used in the low density range $0.07-0.20~$fm$^{-3}$ for cold Pure Neutron Matter (PNM). It should be noted that $\chi$EFT calculations have recently been extended to finite temperature asymmetric nuclear systems \cite{Keller_Wellenhofer_2021prc,Keller_Hebeler_2023prl}. These have also been applied recently to impose constraints on finite temperature EoSs~\cite{Alford2023}. However as these calculations deal with isothermal (fixed temperature) systems, these results have not been implemented in this work as it focuses on isentropic (fixed entropy per baryon) cases.
    \item \textbf{Astrophysical Observations (Astro):} These are appropriate for the high density part of  Asymmetric Nuclear Matter (ANM) and $\beta$ equilibrated EoS. We impose astrophysical constraints in which the maximum mass supported by a cold neutron star has a lower limit of 2.01 $M_{\odot}$, compatible with the current data. GW170817 observation \cite{LSC_2017_GW170817} puts an upper limit on effective tidal deformability ($\tilde{\Lambda}$) to be 720 \cite{Tong_Zhao_2020prc} and radius of a 1.4 M$_{\odot}$ neutron star to be 13.5 km \cite{Most_Weih_2018prl,Eemeli_Tyler_2018prl}.
    \item \textbf{Heavy Ion Collisions (HIC):} These constraints have been derived from heavy-ion experiments at GSI in Germany, namely KaoS \cite{Hartnack_Oeschler_2006prl}, FOPI \cite{Fèvre_Leifels_2016npa} and ASY-EOS \cite{Russotto_2016prc}, which leads to constraints at intermediate densities in the range 1-3$n_{sat}$. The KaoS experiment can be used to put upper limits on nucleon potential ($U_N$) for ANM in the density range 2-3$n_{sat}$. The FOPI experiment can be used to constrain binding energy per nucleon ($E/A$) of Symmetric Nuclear Matter (SNM) in the density range 1.4-2$n_{sat}$, while ASY-EOS results put bounds on the Symmetry Energy of ANM in the supra-saturation density range 1.1-2$n_{sat}$.
\end{enumerate}

\begin{center} 
\captionof{table}{Nuclear saturation parameters considered in this work~\cite{Ghosh_Pradhan_2022_fspas}: The top row is the fixed set of the nuclear parameters used in figures~\ref{fig:T_vs_nb},~\ref{fig:m_vs_r} and~\ref{fig:f/p1_vs_m} and the bottom row shows the range of varied parameters used in the rest of this study.} \label{tab:prior} 
\begin{tabular}{ |c|c|c|c|c|c|c| } 
\hline
&$n_{sat}$ & $E_{sat}$ & $K_{sat}$ & $J_{sym}$ & $L_{sym}$ & $m^*/m$ \\
&($\rm{fm^{-3}}$) & (MeV) & (MeV) & (MeV) & (MeV) & \\
\hline
Fixed & 0.15 & -16.0 & 240 & 32 & 60 & .65 \\
\hline
Varied & 0.14 - & -16.2 - & 200 - & 28 - & 40 - & 0.55 - \\ 
       & 0.17 & -15.8 & 300 & 34 & 70 & 0.75 \\
\hline
\end{tabular}
\end{center}

We provide the uncertainty ranges in nuclear saturation parameters from state-of-the-art experimental data in Table \ref{tab:prior} As in~\cite{Ghosh_Chatterjee_2022epja,Ghosh_Pradhan_2022_fspas}, a hard cut-off filter scheme has been used, in which nuclear saturation properties are varied within their allowed uncertainty range to generate a flat prior and only the results which are within the permitted uncertainty region of the above mentioned constraints are accepted. Imposing these constraints reduces the parameter space of nuclear saturation properties and in turn also of the couplings used in constructing the NL-RMF Lagrangian \ref{eq:NL_RMF}.

\begin{comment}
    
\end{comment}

%%%%%%%%%%%%%%%%%%%%%%%%%%%%%%%%%%%%%%%
\section{Results}
\label{sec:results}

The composition as well as thermodynamical properties of the neutron star change with the evolving temperature during the PNS formation or neutron star merger. In PNS early evolution the star goes through different stages in which the core evolves from a low temperature (low entropy state) and gets heated up (high entropy) with a decrease in charge fraction~\cite{Thapa_Beznogov_2023prd,Prakash_Bombaci_1997phys-rep,Pons_Reddy_1999apj}. It then starts cooling and eventually becomes cold and attains $\beta$ equilibrium. In case of BNS mergers a recent simulation has shown that the entropy per baryon can reach upto values of 3~\cite{Most_Motornenko_2023prd}.
To investigate the effects of temperature and out-of-$\beta$ equilibrium, in this work we consider two thermal configurations, $N(S/A = 1, Y_Q = 0.4)$ and $N(S/A = 2, Y_Q = 0.2)$ representing two different stages of evolution of a hot PNS. The former state corresponds to a hot lepton/electron rich configuration while the latter corresponds to a hotter core with deleptonized configuration~\cite{Thapa_Beznogov_2023prd}. 
We also show comparison with cold $\beta$ equilibrated state in our study.

\subsection{Equation of State}
\label{subsec:EoS_res}
%\nb{Discussion on HIC constraints are to be written}\\
We investigate the effect of non-zero temperature and out-of-$\beta$ equilibrium condition on the macroscopic NS properties. The results after imposing constraints from $\chi$EFT, Astro and HIC constraints are shown for the thermal case $N(2, 0.2)$ in fig.~\ref{fig:N(2, 0.2)_eos}. The bands of EoSs corresponds to different sets of nuclear saturation properties consistent with filters mentioned above. The imposition of Astro and HIC constraints eliminates certain stiffer EoSs from the parameter set as evident from the figure.
\begin{figure}[H]
    \centering
    \includegraphics[width=\linewidth]{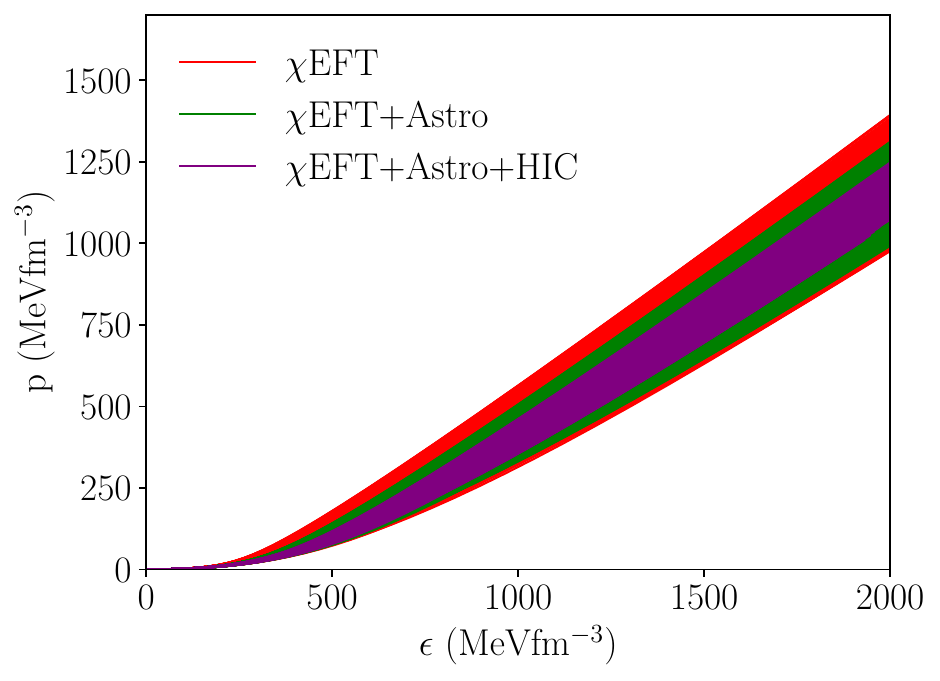}
    \caption{EoS bands for $N(2, 0.2)$ generated after imposing $\chi$EFT, Astro and HIC constraints on nuclear saturation properties.}
    \label{fig:N(2, 0.2)_eos}
\end{figure}

\begin{figure}[H]
    \centering
    \includegraphics[width=\linewidth]{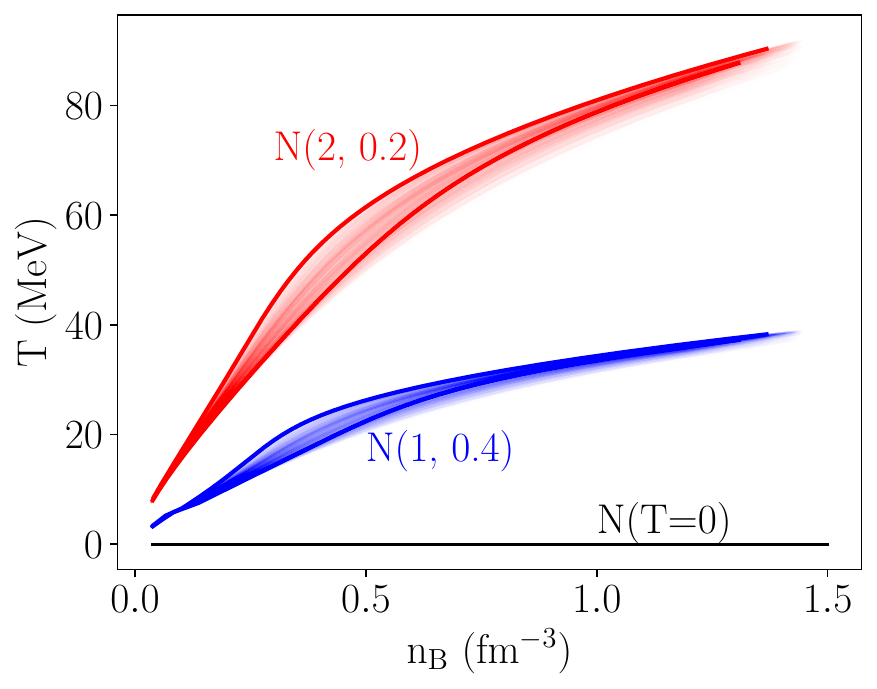}
    \caption{Core temperature profiles obtained on imposing $\chi$EFT and Astro constraints.}
    \label{fig:T_nb_combined}
\end{figure}

\begin{figure}[H]
    \centering
    \includegraphics[width=\linewidth]{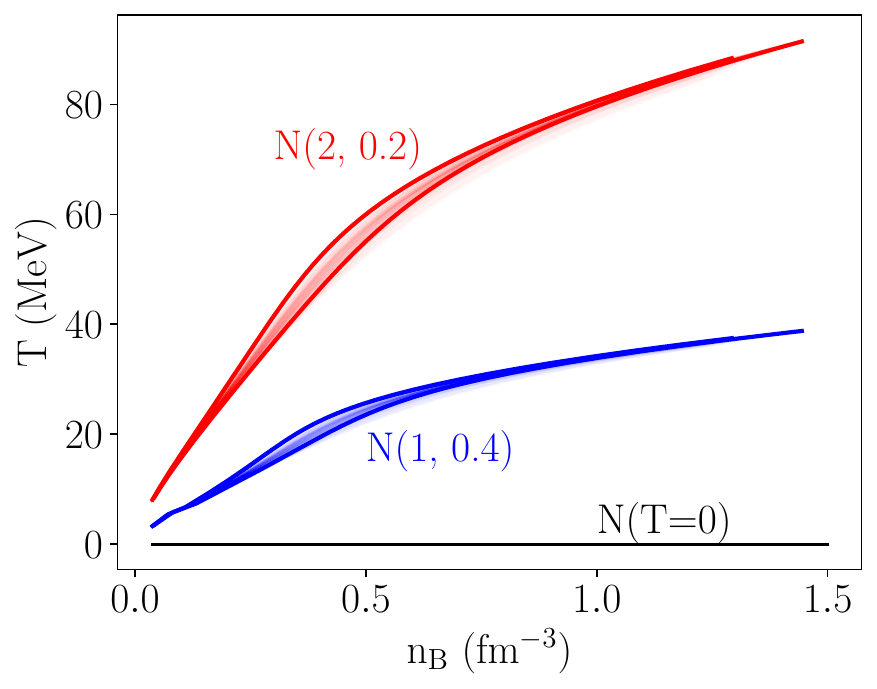}
    \caption{Core temperature profiles obtained on imposing $\chi$EFT + Astro + HIC constraints.}
    \label{fig:T_nb_combined_HIC}
\end{figure}

The temperature profiles in the core of these stars using  the $\chi$EFT + Astro constraints are shown in fig.~\ref{fig:T_nb_combined}. For low density ($n_B<0.1$ fm$^{-3}$) there is no significant variation in the temperature profiles for the thermal cases. However, saturation properties do induce appreciable changes in temperature at intermediate densities. For $N(1, 0.4)$ the variation is limited to about $n_B<0.9$ fm$^{-3}$ while that of $N(2, 0.2)$ extends to the entire density range considered. fig.~\ref{fig:T_nb_combined_HIC} shows temperature profiles after inclusion of HIC filter and we see that temperature spreads across intermediate and high densities have reduced. However qualitative difference of spreads between $N(1, 0.4)$ and $N(2, 0.2)$ across densities still persists.

\begin{comment}
\nb{We observe that at higher densities the temperature profiles get flattened. This is more prominent in case of $N(1, 0.4)$ case as there are higher nucleation of positrons at high densities due to higher charge fraction. The new anti-particle degrees of freedom contribute to the overall entropy and hence temperature do not increase rapidly with increasing density.} \dc{check. Compared to which work?}
\end{comment}

\subsection{Mass-Radius Relations}\label{subsec:M-R_res}
%\nb{Discussion on HIC constraints are to be written}\\
The Mass-Radius relations for $N(2, 0.2)$ are shown in fig.~\ref{fig:N(2, 0.2)_mr}. The filter imposed by Astro constraints on stiffer EoSs translates to lower values of maximum masses and radii supported by the configurations. We will see later on in subsection~\ref{subsec:correlations} that the maximum mass is strongly anti-correlated with effective nucleon mass ($m^*/m$) and therefore this parameter controls the stiffness of the EoS. It also strongly effects other macroscopic properties such as radii and tidal deformability. 
\begin{figure}[H]
    \centering
    \includegraphics[width=\linewidth]{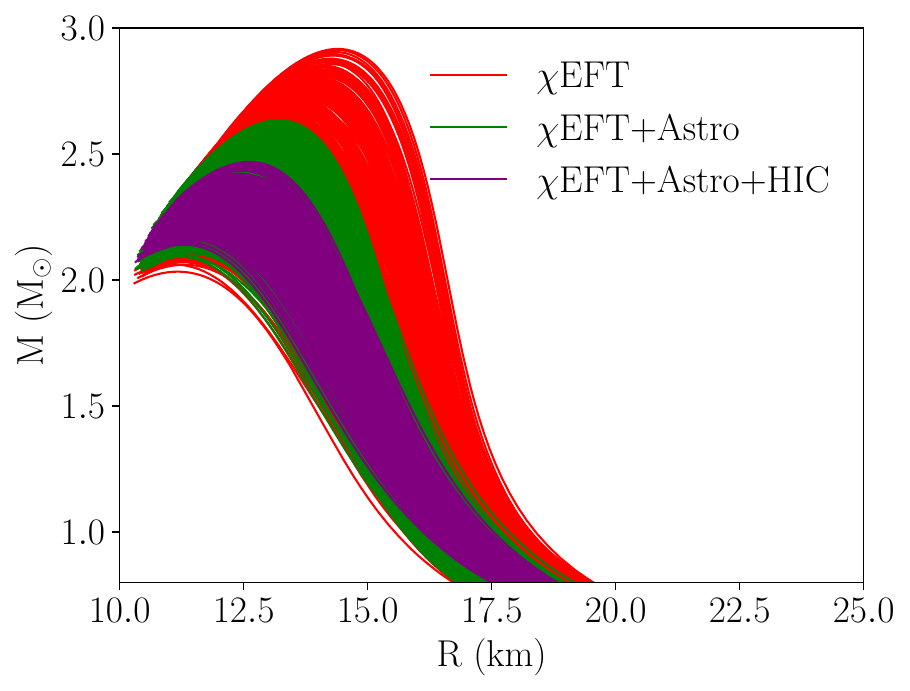}
    \caption{Mass-Radius posterior plots for $N(2, 0.2)$. Constraints are same as in fig.~\ref{fig:N(2, 0.2)_eos}. We observe that maximum mass supported after application of different filters decreases.}
    \label{fig:N(2, 0.2)_mr}
\end{figure}
\begin{figure}[H]
    \centering
    \includegraphics[width=\linewidth]{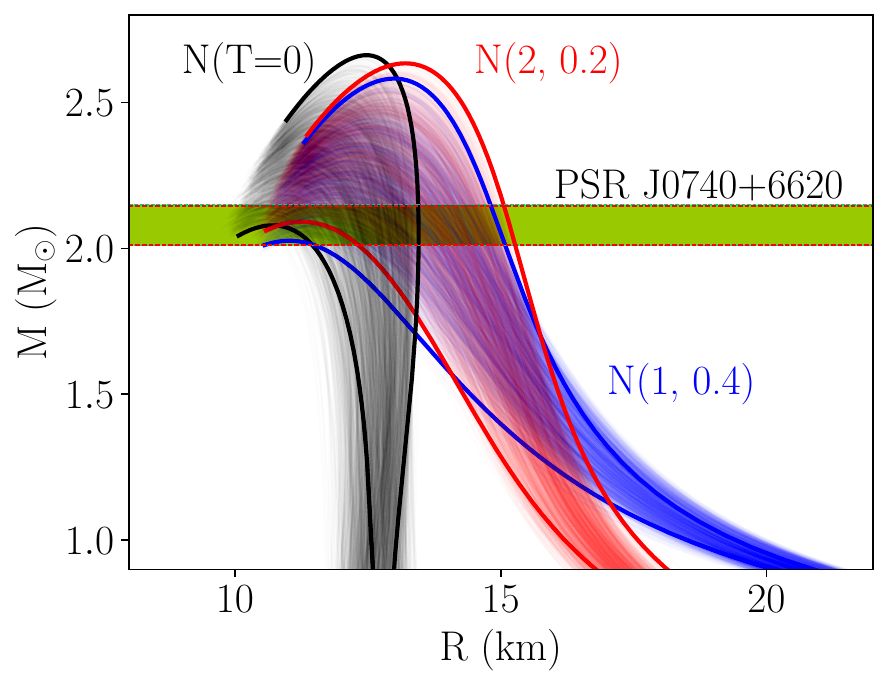}
    \caption{Mass-Radius configuration for both the thermal cases along with $N(T=0)$ $\beta$ equilibrated matter. The constraints used are $\chi$EFT + Astro. The solid lines represent the stiffest and softest configurations for each case.}
    \label{fig:mr_combined}
\end{figure}
We also display these results for the thermal case $N(1, 0.4)$ and their comparison with cold $\beta$ equilibrated matter in fig.~\ref{fig:mr_combined}. One can see that the cold $\beta$ equilibrated configuration is the most compact one and the highest maximum mass it supports  is almost identical to $N(2, 0.2)$. The $N(2, 0.2)$ configuration is more compact for higher masses while for $N(1, 0.4)$ this is the case for low and intermediate mass configurations. After imposition of HIC constraints (fig. \ref{fig:mr_combined_HIC}), the maximum masses supported by each thermal configuration further reduce. Most notably $M_{max}>2.5M_{\odot}$ configurations are eliminated for all three cases. We discuss in subsection \ref{subsec:correlations} how the imposition of these constraints affect masses, radii and tidal deformability.

\begin{figure}[H]
    \centering
    \includegraphics[width=\linewidth]{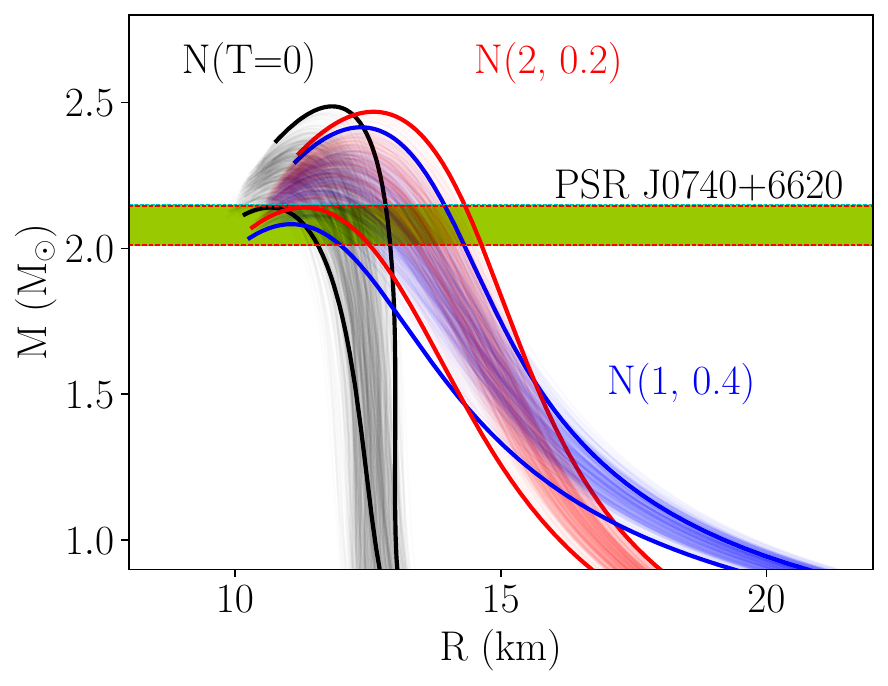}
    \caption{Same as fig. \ref{fig:mr_combined}. The constraints used are $\chi$EFT + Astro + HIC.}
    \label{fig:mr_combined_HIC}
\end{figure}

\subsection{\texorpdfstring{$f$}{Lg}-mode frequency and damping time}\label{subsec:f-mode_res}
\begin{figure}[H]
    \centering
    \includegraphics[width=\linewidth]{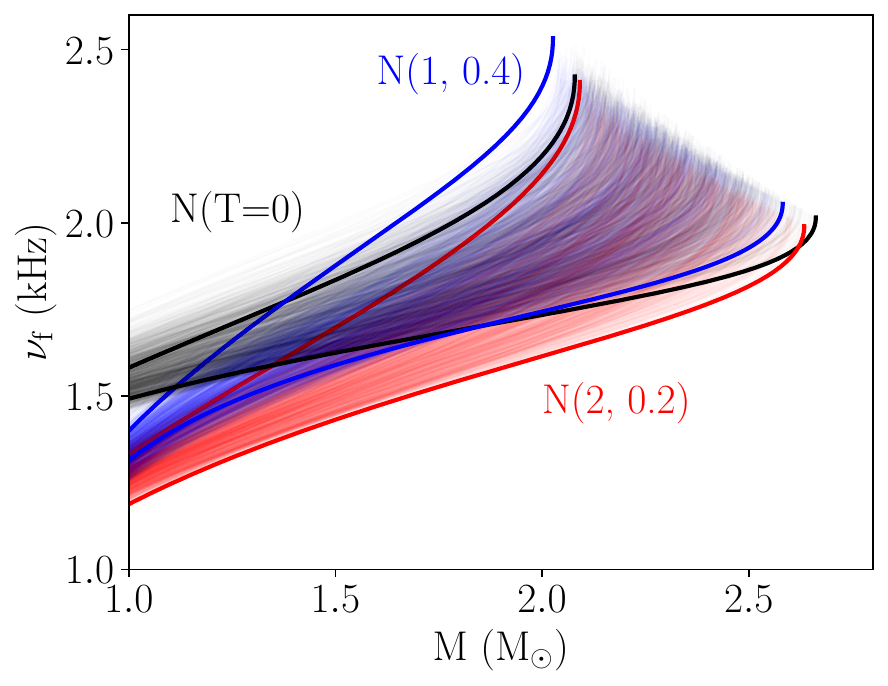}
    \caption{$f$-mode frequencies with $\chi$EFT + Astro constraints.}
    \label{fig:fr_m_combined}
\end{figure}

%\nb{Discussion on HIC constraints are to be written}\\
As discussed in subsection~\ref{subsec:modes}, after imposing the constraints mentioned above, the $f$-mode frequencies and damping times are calculated. The results for the case $N(2, 0.2)$ are shown in fig.~\ref{fig:N(2, 0.2)_fm_GR} and fig.~\ref{fig:N(2, 0.2)_taum_GR}.
\begin{figure}[H]
    \centering
    \includegraphics[width=\linewidth]{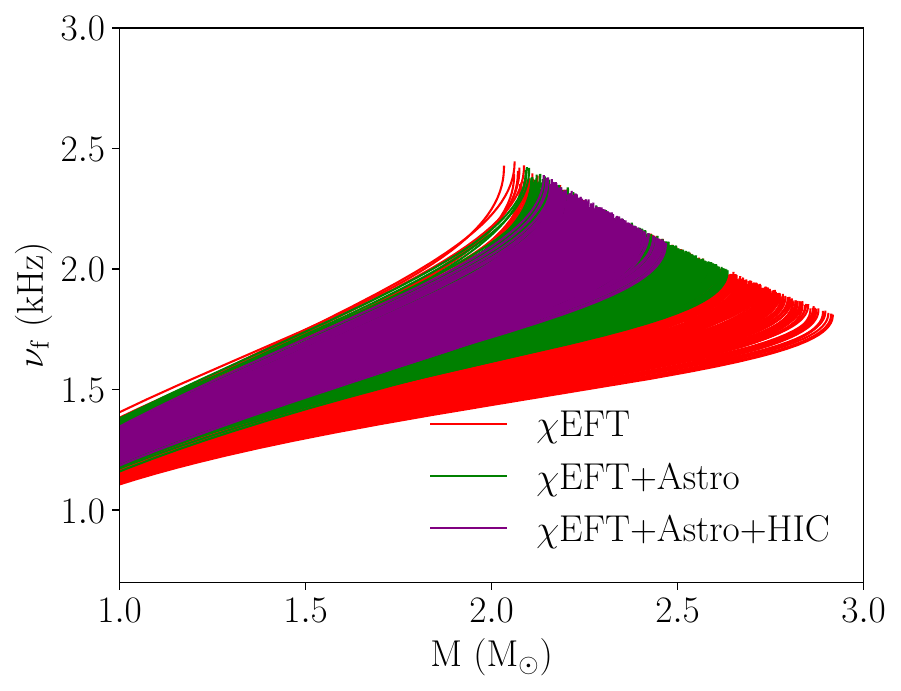}
    \caption{$f$-mode frequency v/s mass posterior plots for $N(2, 0.2)$. Constraints are same as in fig.~\ref{fig:N(2, 0.2)_eos}.}
    \label{fig:N(2, 0.2)_fm_GR}
\end{figure}
\begin{figure}[H]
    \centering
    \includegraphics[width=\linewidth]{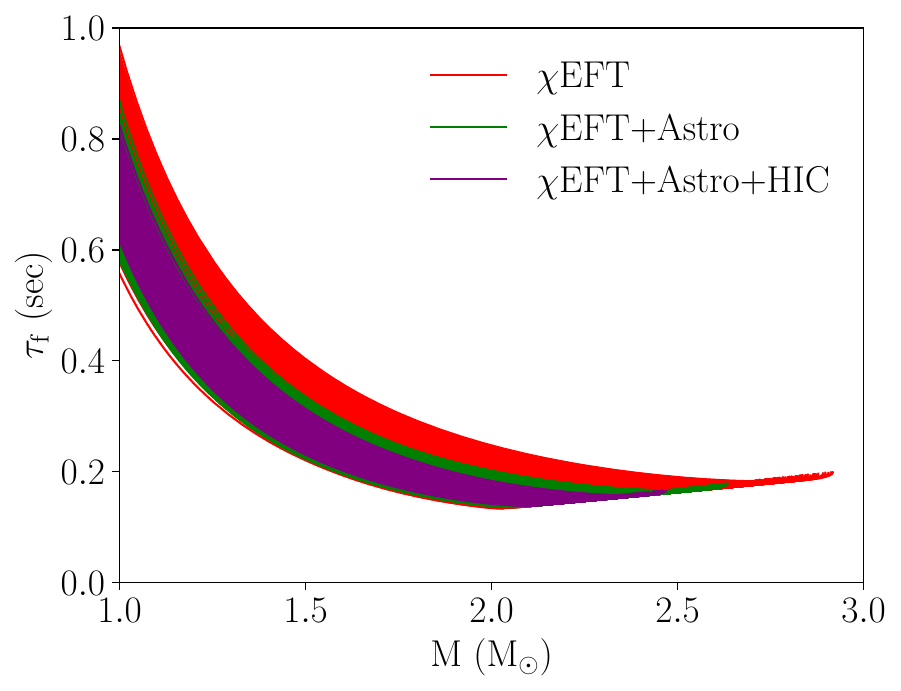}
    \caption{$f$-mode damping time v/s mass posterior plots for $N(2, 0.2)$. Constraints are same as in fig.~\ref{fig:N(2, 0.2)_eos}.}
    \label{fig:N(2, 0.2)_taum_GR}
\end{figure}
Next we show the combined $f$-mode spectrum for all the three cases with $\chi$EFT + Astro constraints in fig.~\ref{fig:fr_m_combined}. From the figure we can see that $f$-mode frequencies are lower for low mass ($M \lesssim1.1$ M$_{\odot}$) NSs for both the thermal cases in comparison with cold and $\beta$ equilibrated stars. For higher masses all the three configurations have overlapping regions similar to their mass-radius relations (fig. \ref{fig:mr_combined}). The effective nucleon mass again plays a vital role in determining $f$-mode frequencies and damping times as have discussed in subsection~\ref{subsec:correlations}. $n_{sat}$ also plays some role in $f$-mode oscillations of intermediate mass configurations (see subsection~\ref{subsec:correlations}).

\begin{comment}
    \dc{For the case $N(1, 0.4)$ the $f$-mode frequencies are highest for $M \gtrsim1.4$ M$_{\odot}$ if the EoS is softer and $M>1.7$ M$_{\odot}$ if it is stiffer. All the $N(2, 0.2)$ states have the lowest $f$-mode frequencies across the entire mass range. ?? DID NOT FOLLOW}
\end{comment}

\begin{figure}[H]
    \centering
    \includegraphics[width=\linewidth]{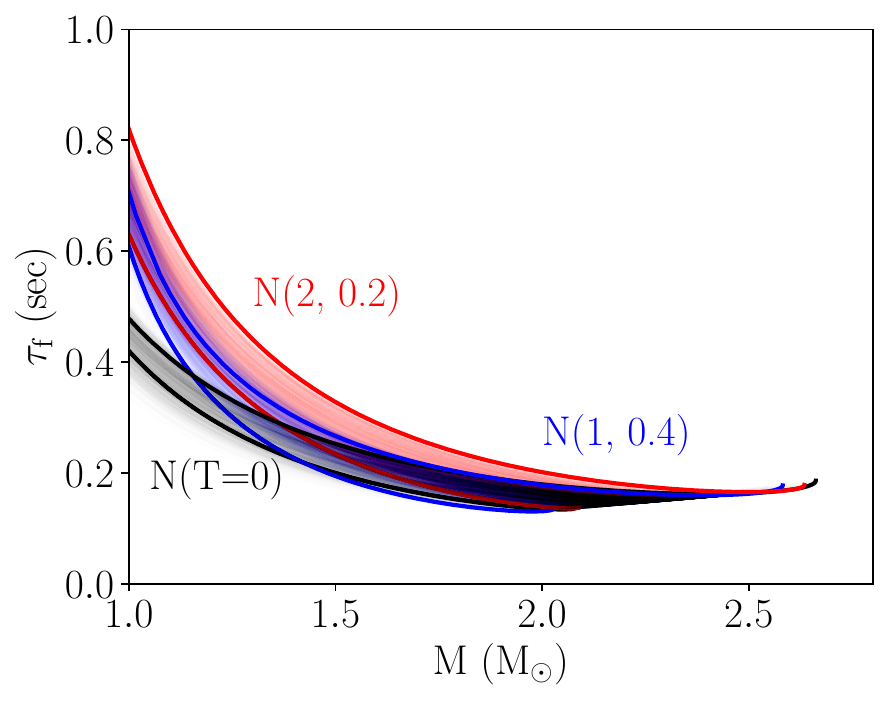}
    \caption{Damping of $f$-modes from $\chi$EFT + Astro constraints. We see overlapping regions in high mass configurations.}
    \label{fig:tauf_m_combined}
\end{figure}

Fig. \ref{fig:tauf_m_combined} shows damping timescales for the three cases. Cold neutron stars have the fastest damping for low mass ($\rm{M\lesssim1.1M_{\odot}}$) cases. $N(2, 0.2)$ configuration have the slowest damping in low and intermediate mass regions. For higher masses we again have overlapping regions. 
\begin{comment}
    \dc{For $N(1, 0.4)$, damping times are almost similar to that of $N(T=0)$ for $M\gtrsim1.7$ M$_{\odot}$ in case of stiffer EoSs and for $M\gtrsim1.4$ M$_{\odot}$ in case of softer EoSs. All the $N(2, 0.2)$ states show the slowest damping across the entire mass range. ?? DID NOT FOLLOW}
\end{comment}

\begin{figure}[H]
    \centering
    \includegraphics[width=\linewidth]{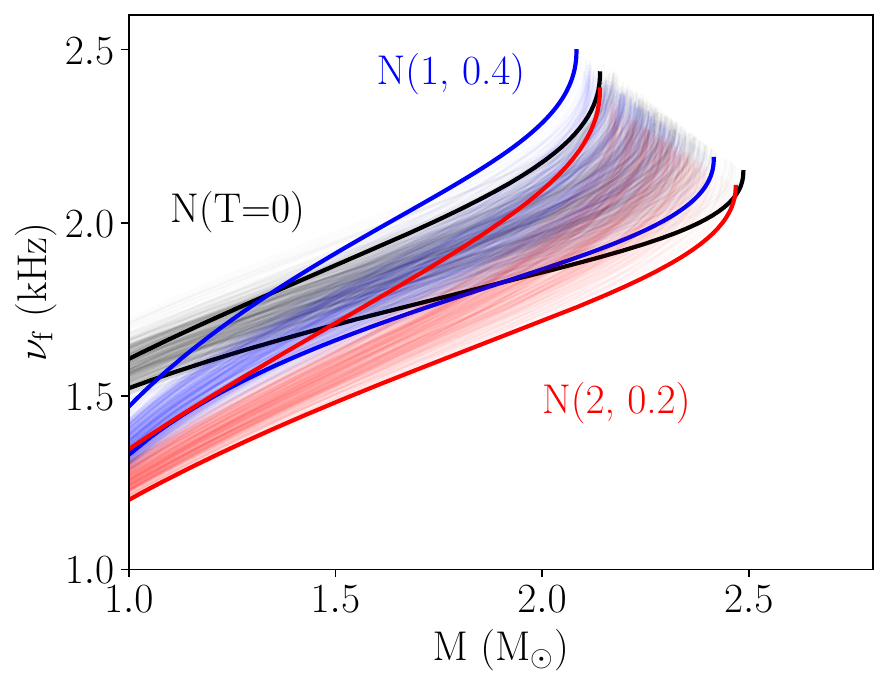}
    \caption{$f$-mode frequencies with $\chi$EFT + Astro + HIC constraints.}
    \label{fig:fr_m_combined_HIC}
\end{figure}

\begin{figure}[H]
    \centering
    \includegraphics[width=\linewidth]{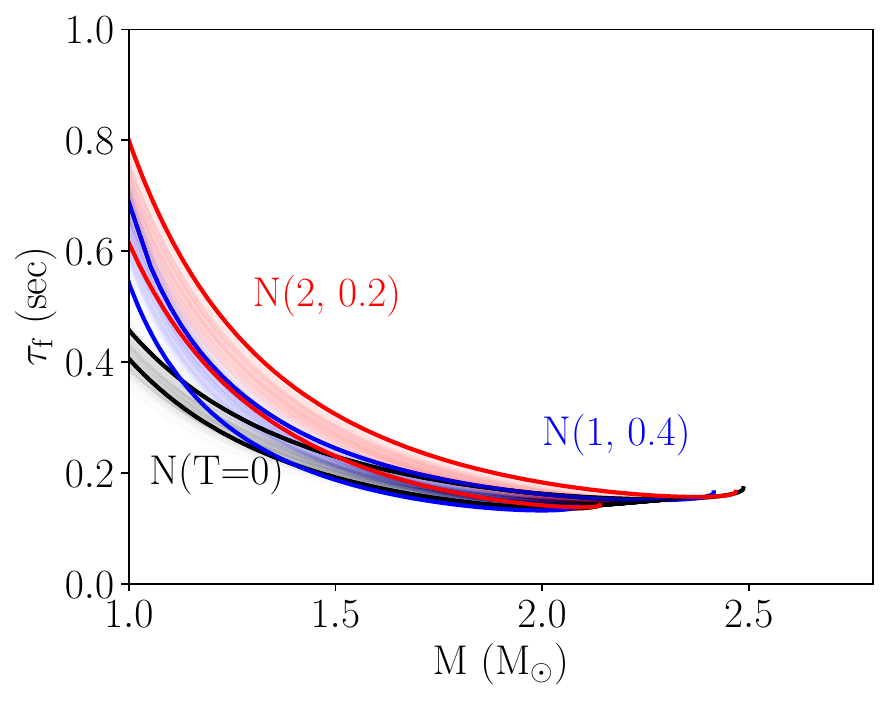}
    \caption{Damping of $f$-modes from $\chi$EFT + Astro + HIC constraints.}
    \label{fig:tauf_m_combined_HIC}
\end{figure}

The $f$-mode frequencies and damping times obtained after applying HIC constraints are shown in figs. \ref{fig:fr_m_combined_HIC} and \ref{fig:tauf_m_combined_HIC}. These constraints eliminate certain regions of figs. \ref{fig:fr_m_combined} and \ref{fig:tauf_m_combined}, however the qualitative behaviour does not change much. 
We see that the frequencies and damping times are inversely correlated for the thermal cases, a fact which is already known for cold NSs. The correlations of $f$-mode frequencies and damping times with nuclear saturation properties are discussed in \ref{subsec:correlations}.

\subsection{Universal relations}
\label{subsec:UR}

%\dc{Please complete the writeup of this section. Please add the correlation study including HIC} \nb{will it be better to put this section after the correlation section \ref{subsec:correlations} ?}\\
In this section, we discuss the effects of nuclear saturation properties on the Universal Relations (URs) of hot NSs. We focus on the two well-studied URs of cold neutron stars, i.e., Love-Compactness ($\Lambda-C$) and $f$-mode frequency-Love ($f-\Lambda$) relations. In addition to the two thermal cases considered earlier, we also show the variation of these URs with a wider range of thermal conditions.
\subsubsection{\texorpdfstring{$C$}{Lg}-Love Relations} \label{subsubsec:Love-C_UR}
\begin{figure}[H]
    \centering
    \includegraphics[width=\linewidth]{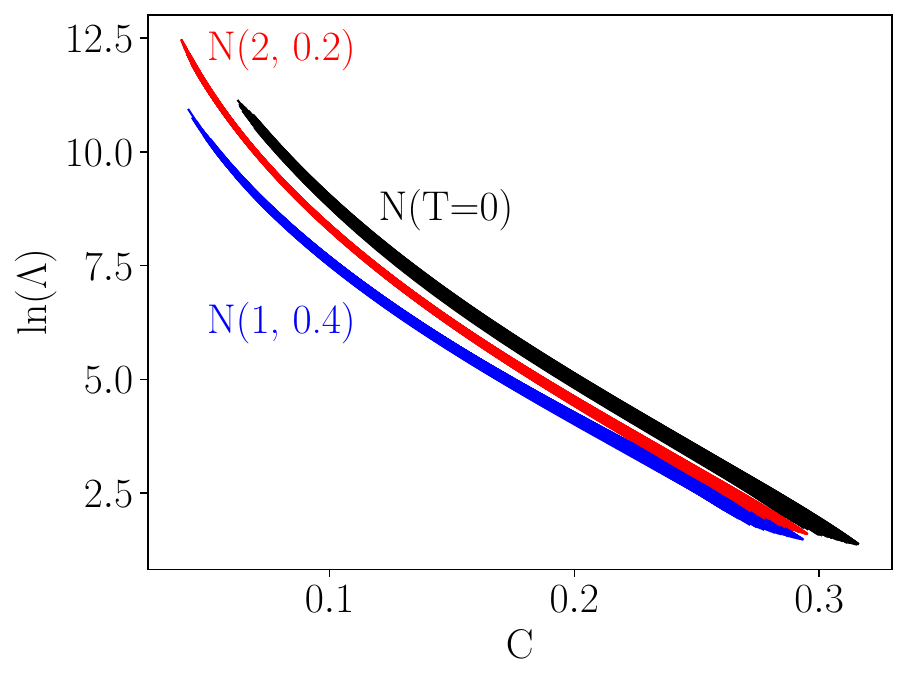}
    \caption{$\Lambda~v/s~C$ universality for the three cases.}
    \label{fig:Lambda_C_combined}
\end{figure}

The EoS independent UR between dimensionless tidal deformability ($\Lambda$) and compactness ($C$) was studied in Maselli~et~al., where they provide the fitting coefficients in equation (4) of their paper \cite{Maselli_Cardoso_2013prd}. We display ${\ln(\Lambda)-C}$ plots from EoSs generated using constraints from $\chi$EFT + Astro in fig. \ref{fig:Lambda_C_combined}. We see that like cold $\beta$ equilibrated stars $(N(T=0))$, the plots for two thermal cases also fall in their respective curves which are distinct from that of $N(T=0)$ curve, indicating that these configurations do not follow the same URs. The deviations from the cold case are more for higher values of $\Lambda$. We also see that compactness corresponding to a particular $\Lambda$ is highest for $N(T=0)$ followed by $N(2, 0.2)$ and $N(1, 0.4)$ indicating the order in compactness in the three configurations. The $C-\rm{ln}(\Lambda)$ polynomial fitting can be given as - 

\begin{equation} \label{eq:C-Lambda}
    C = \sum^2_{i=0} a_i [\rm{ln(\Lambda)}]^i~.
\end{equation}

\begin{center} 
\captionof{table}{Fit relations for $C-\rm{ln}(\Lambda)$ UR for the two thermal configurations}
\label{tab:C-Lambda_UR}
\begin{tabular}{ |c|c|c|c| } 
\hline
 & $a_0$ & $a_1$ & $a_2$  \\
\hline
\hline
$N(1, 0.4)$ & $3.556\times10^{-1}$ & $-4.209\times10^{-2}$ & $1.173\times10^{-3}$  \\ 
\hline
\hline
$N(2, 0.2)$ & $3.598\times10^{-1}$ & $-4.070\times10^{-2}$ & $1.182\times10^{-2}$  \\ 
\hline
\end{tabular}
\end{center}

To further investigate how the thermal effects break this universality, we separately vary entropy per baryon ($S/A$) and charge fractions ($Y_Q$) keeping nuclear saturation properties fixed. First we show the variation on changing $S/A$ from $1-2$ by fixing $Y_Q=0.4$, and in $\beta$ equilibrium. Next we fix $S/A=1$ and vary $Y_Q$ in the range $0.01-0.40$. 

\begin{figure}[H]
    \centering
    \includegraphics[width=\linewidth]{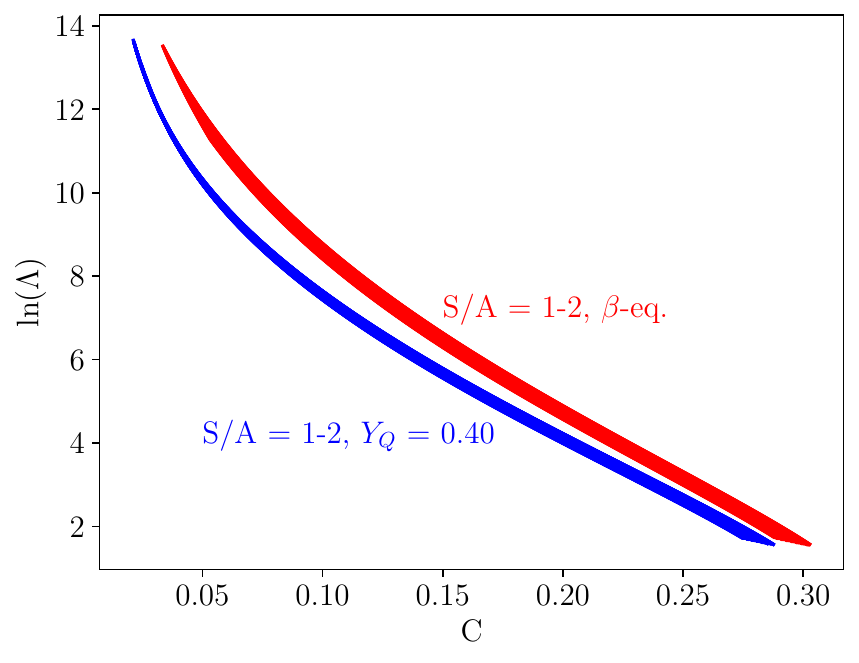}
    \caption{Effect of $S/A$ on $\Lambda-C$ universality. Nuclear saturation parameters are fixed to same values as in fig. \ref{fig:T_vs_nb}. $S/A$ has been varied between $1-2$ in steps of 0.01.}
    \label{fig:Lambda_C_s_variation}
\end{figure}

\begin{figure}[H]
    \centering
    \includegraphics[width=\linewidth]{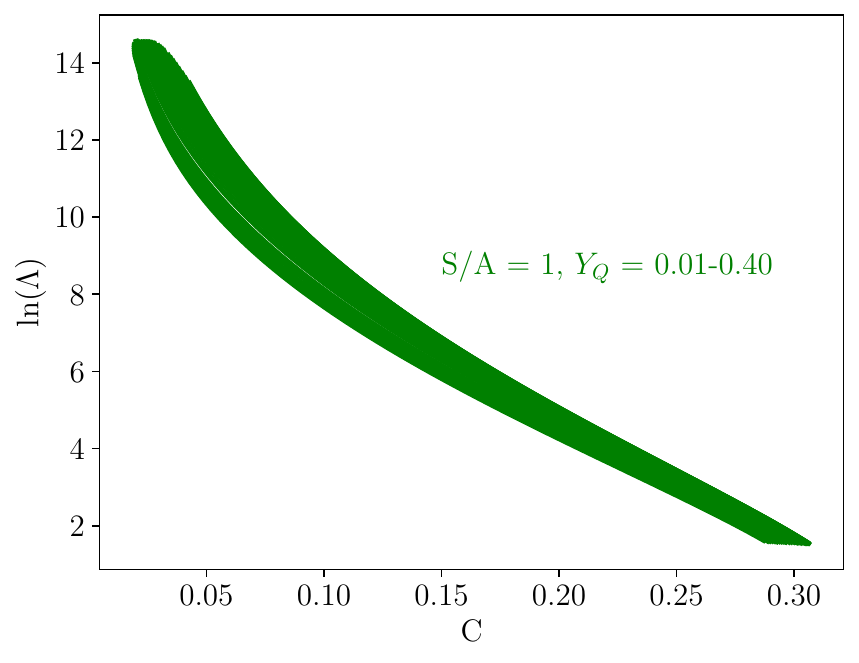}
    \caption{Effect of $Y_Q$ on $\Lambda-C$ universality. Nuclear saturation properties are fixed as in fig. \ref{fig:T_vs_nb}. $Y_Q$ has been varied between 0.01-0.40 in steps of 0.01.}
    \label{fig:Lambda_C_y_variation}
\end{figure}

It is clear that $\Lambda-C$ relation depends not only on charge fraction but also on whether the matter is in $\beta$ equilibrium when we vary $S/A$. The spread in universality in fig. \ref{fig:Lambda_C_s_variation} is much less than that of fig. \ref{fig:Lambda_C_y_variation}. From the results in fig.~\ref{fig:Lambda_C_s_variation} one can obtain universal fits similar to the previous works ~\cite{Maselli_Cardoso_2013prd,PradhanPRD}. We provide the universal fits for the two thermal configurations in Table \ref{tab:C-Lambda_UR}. On the other hand, we see from fig. \ref{fig:Lambda_C_y_variation} that $Y_Q$, which is a measure of out-of-$\beta$ equilibrium, is mainly responsible for breaking this universality and there is no possibility to obtain a fit relation.

\subsubsection{\texorpdfstring{$f$}{Lg}-Love Relations} \label{subsubsec:f-Love_UR}
It is known that for cold NSs $f$-mode frequencies scale with certain combinations of astrophysical observables, such as $M$, $R$ and $\Lambda$. It was shown that $f$-mode frequency has an approximately linear relation with square root of average density ($\sqrt{M/R^3}$) while the normalized damping time ($\tau_f R^4/M^3$) scales with stellar compactness \cite{Andersson_Kokkotas_1998mnras}. These fit relations are found in the works of Andersson and Kokkotas \cite{Andersson_Kokkotas_1998mnras}, Benhar and Ferrari \cite{Benhar_Ferrari_2004prd}, Doneva et al. \cite{Doneva_Gaertig_2013prd}, Pradhan and Chatterjee \cite{Pradhan_Chatterjee_2021prc}.

However, it turns out that these fits are composition dependent and have large deviations depending on the EoSs considered. Pradhan~et~al. \cite{Pradhan_Chatterjee_2022prc} also provided other improved universal fits which correlate mass-scaled frequencies ($M \omega$) (real and imaginary) with logarithm of dimensionless tidal deformability ($\ln(\Lambda)$). This polynomial fitting can be given as - 
\begin{equation}
    M\omega = \sum^5_{i=0} \alpha_i[\rm{ln(\Lambda)}]^i~.
\end{equation}

\begin{figure}[H]
    \centering
    \begin{subfigure}[b]{0.44\textwidth}
        \includegraphics[width=\linewidth]{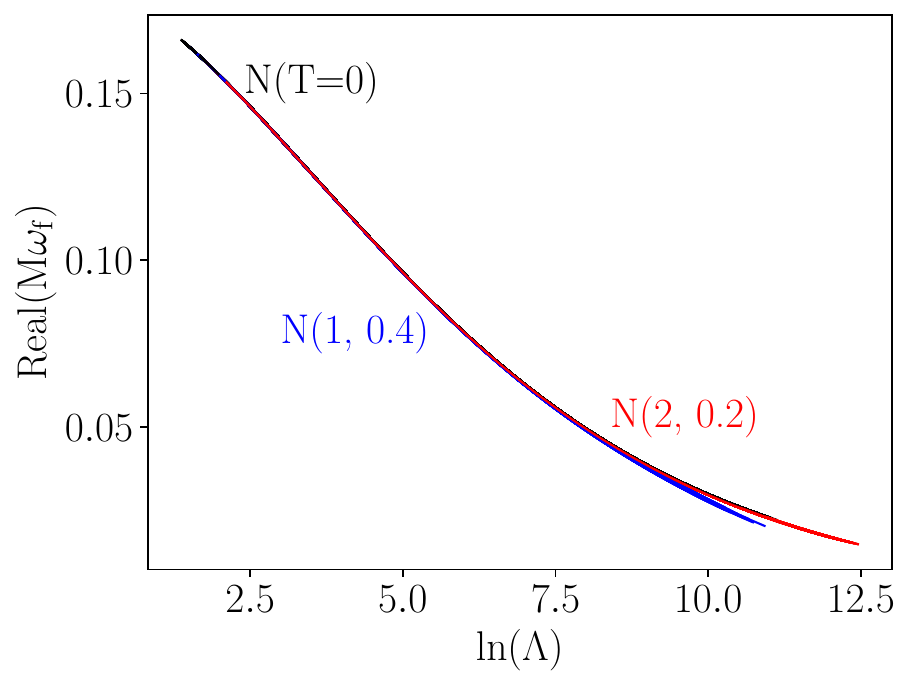}
        \caption{Real($M\omega)~v/s~\rm{ln(\Lambda)}$}
        \label{fig:mwr_lambda_combined}
    \end{subfigure}
    %\hfill
    \begin{subfigure}[b]{0.41\textwidth}
        \includegraphics[width=\linewidth]{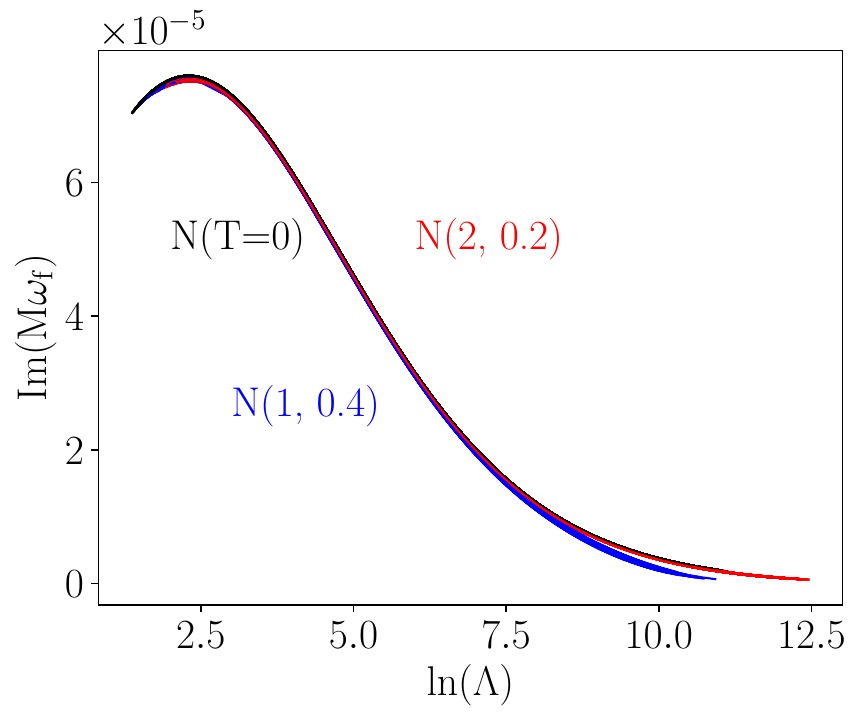}
        \caption{Im($M\omega)~v/s~\rm{ln(\Lambda)}$}
        \label{fig:mwi_lambda_combined}
    \end{subfigure}
    \caption{$f$-Love universality for the three cases. The relations closely resembles $N(T=0)$ state and deviations are seen only for high $\Lambda$ or consequently less compact and less massive configurations.}
    \label{fig:f-Love}
\end{figure}

Here we examine this $f$-Love universality in the context of hot NSs. The $f$-Love relations after imposing constraints from $\chi$EFT and Astro are shown in figs. \ref{fig:mwr_lambda_combined} and \ref{fig:mwi_lambda_combined}. Unlike fig. \ref{fig:Lambda_C_combined}, we observe negligible spreads for both thermal cases. Small deviations from the universality appear for NSs with $\Lambda>e^8\sim3000$ or with masses close to or less than solar mass. After the post-bounce stage, a PNS becomes more and more compact as it evolves, consequently $\Lambda$ goes from high to low and the star approaches a cold $\beta$ equilibrated state. The $f$-Love fit coefficients for the two thermal configurations are given in Table~\ref{tab:f-Love_UR}.

These findings are reasonable, as at higher $\Lambda$ the star can be assumed to be in the $N(1, 0.4)$ representative state and this configuration shows the highest deviation from cold $f$-Love universality. The $N(2, 0.2)$ state is representative of intermediate $\Lambda$ and it has close resemblance with the cold $\beta$ equilibrated configuration. The close resemblance these URs have with $N(T=0)$ promotes us to study them across ranges of $S/A$ and $Y_Q$. The conclusions are somewhat different to those for $C$-Love relations (see Appendix sections~\ref{appendix:f-Love_entropy} and~\ref{appendix:f-Love_yq}).

%\onecolumngrid

\onecolumngrid
\begin{center} 
\captionof{table}{Fit relations for $f$-Love UR for the two thermal configurations} \label{tab:f-Love_UR}
\begin{tabular}{ |c|c|c|c|c|c|c| } 
\hline
& $\alpha_0$ & $\alpha_1$ & $\alpha_2$ & $\alpha_3$ & $\alpha_4$ & $\alpha_5$ \\
\hline
\hline
$N(1, 0.4)$ & $1.812\times10^{-1}$ & $-5.435\times10^{-3}$ & $-4.999\times10^{-3}$ & $7.106\times10^{-4}$ & $-3.798\times10^{-5}$ & $7.034\times10^{-7}$ \\ 
& $+i2.237\times10^{-5}$ & $+i5.599\times10^{-5}$ & $-i1.863\times10^{-5}$ & $+i2.142\times10^{-6}$ & $-i1.004\times10^{-7}$ & $+i1.452\times10^{-9}$ \\
\hline
\hline
$N(2, 0.2)$ & $1.809\times10^{-1}$ & $-5.110\times10^{-3}$ & $-5.141\times10^{-3}$ & $7.405\times10^{-4}$ & $-4.114\times10^{-5}$ & $8.486\times10^{-7}$ \\ 
& $+i1.411\times10^{-5}$ & $+i6.519\times10^{-5}$ & $-i2.239\times10^{-5}$ & $+i2.849\times10^{-6}$ & $-i1.619\times10^{-7}$ & $+i3.466\times10^{-9}$ \\
\hline
\end{tabular}
\end{center}

%\twocolumngrid

\begin{comment}
\begin{figure}[H]
    \centering
    \includegraphics[width=\linewidth]{mwr_lambda_combined.pdf}
    \caption{Caption}
    \label{fig:mwr_lambda_combined}
\end{figure}

\begin{figure}[H]
    \centering
    \includegraphics[width=\linewidth]{mwi_lambda_combined.pdf}
    \caption{Caption}
    \label{fig:mwi_lambda_combined}
\end{figure}
\end{comment}
%\begin{comment}

%\end{comment}

\begin{comment}
    \dc{This part may be moved to the Appendix.} \nb{We should keep $\rm{Real(M\omega_f)~v/s~ln(\Lambda)}$ figs. \ref{fig:Real(Mw)_Lambda_s_variation} and \ref{fig:Real(Mw)_Lambda_y_variation} and the discussions. We can remove figs. \ref{fig:Im(Mw)_Lambda_s_variation} and \ref{fig:Im(Mw)_Lambda_y_variation} as they point to similar conclusions.}
\end{comment}

%\newpage
\twocolumngrid
\subsection{Correlation studies}
\label{subsec:correlations}

From the correlation studies in the case of cold $\beta$ equilibrated matter, it was found that the most important nuclear saturation parameter which affects the NS observable properties is the nucleon effective mass $m^*$ and to a certain extent the nuclear saturation density $n_{sat}$ ~\cite{Ghosh_Chatterjee_2022epja,Pradhan_Chatterjee_2022prc}. In this section we will probe possible correlations of the nuclear parameters with NS observables and also among each for the thermal cases. First we study the effect of imposing the $\chi$EFT + Astro constraints on the correlations. Next, we also include the HIC constraints, and investigate how the correlations get altered. As the HIC results are known to be dependent on the transport models, we have treated them separately with caution (see~\cite{Ghosh_Chatterjee_2022epja} for a more detailed discussion). \\
\begin{comment}
    For this analysis we use the Pearson’s linear correlation coefficient. For two random variables $A$ and $B$, it is given by:
\begin{equation}\label{eq:Pearson}
    R(A,B) = \frac{Cov(A,B)}{S(A) \times S(B)}
\end{equation}
where $Cov(A,B)$ is the covariance of two random variables and $S(A)$ and $S(B)$ are standard deviations of $A$ and $B$ variables respectively.
\end{comment}

\subsubsection{\texorpdfstring{$\chi$}{Lg}EFT + Astro}
The correlations among the nuclear saturation parameters and the NS observable properties on imposing $\chi$EFT + Astro constraints are shown for the two thermal cases in figs. \ref{fig:correlation_N(1, 0.4)} and \ref{fig:correlation_N(2, 0.2)}. We describe the findings in detail below. 

\begin{enumerate}
    \item \textbf{Correlations among nuclear saturation properties:} 
    \begin{itemize}
        \item Effective nucleon mass $m^*$ shows moderate anti-correlation with $n_{sat}$.
        \item Relatively strong correlation is observed between $J_{sym}$ and $L_{sym}$. This is consistent with previous studies \cite{Ghosh_Chatterjee_2022epja}.
        \item $E_{sat}$ and $K_{sat}$ do not show correlations with other saturation properties.
    \end{itemize}
    
    \item \textbf{Correlations of nuclear saturation properties with NS observables:} 
    \begin{itemize}
        \item For both thermal cases, $n_{sat}$ shows moderate correlations/anti-correlations with properties ($R$, $\Lambda$, $\nu_f$, $\tau_f$) of canonical $\rm{1.4M_{\odot}}$ hot NSs.
        \item On the other hand $n_{sat}$ shows weak correlations/anti-correlations with properties ($R$, $\Lambda$, $\nu_f$, $\tau_f$) of canonical $\rm{2M_{\odot}}$ hot NSs and it does not have any effect on maximum mass supported by the corresponding  EoS.
        \item Effective nucleon mass $m^*$ has strong anti-correlation with $M_{max}$, it also shows relatively stronger correlations/anti-correlations with $R_{\rm{2M_{\odot}}}$, $\Lambda_{\rm{2M_{\odot}}}$, $\nu_{f, {\rm{2M_{\odot}}}}$ and $\tau_{f, {\rm{2M_{\odot}}}}$ for both cases.
       % \item For both cases effective mass has relatively strong correlations with observables of  $\rm{2M_{\odot}}$ 
        %\sout{except $\rm{T_{c,2M_{\odot}}}$ with which it has a moderate correlation}. \dc{Please remove core temperature from observables and also correlation matrices.} 
        \item However for $\rm{1.4M_{\odot}}$ hot NSs, $m^*/m$ has weak correlations/anti-correlations with the observables for the case $N(1, 0.4)$ but moderate correlations/anti-correlations for $N(2, 0.2)$ observables. Previous study with cold NSs have also shown moderate correlations of  $m^*/m$ with $\rm{1.4M_{\odot}}$ observables \cite{Ghosh_Chatterjee_2022epja}. We can conclude that during the early phase of hot NS with low entropy and high lepton fraction, $m^*/m$ have lesser effect on properties of intermediate mass configurations.
        %\item \sout{$K_{sat}$ has weaker to negligible correlations/anti-correlations with all observables except $\rm{T_{c,1.4M_{\odot}}}$. It indicates that $K_{sat}$ plays a role in setting up thermodynamic condition inside intermediate mass PNSs.}
    \end{itemize}

    \item \textbf{Correlations among NS observables:}
    \begin{itemize}
        \item Maximum mass is strongly correlated/anti-correlated with properties of canonical $\rm{2M_{\odot}}$ observables for both cases.
        \item With observables of $\rm{1.4M_{\odot}}$, these correlations are moderate for the state $N(1, 0.4)$ and strong for the case $N(2, 0.2)$. %\sout{$\rm{T_{c,1.4M_{\odot}}}$ however shows no correlation with $\rm{M_{max}}$ in both cases.}
       % \item Correlation between $\rm{T_{c,1.4M_{\odot}}}$ and $\rm{T_{c,2M_{\odot}}}$ is stronger for the case N(1, 0.4). $\rm{T_{c,2M_{\odot}}}$ show strong correlations/anti-correlations with observables of both $\rm{1.4M_{\odot}}$ and $\rm{2M_{\odot}}$. $\rm{T_{c,1.4M_{\odot}}}$ have moderate correlations/anti-correlations with observables of $\rm{1.4M_{\odot}}$. For observables of $\rm{2M_{\odot}}$ PNSs, $\rm{T_{c,1.4M_{\odot}}}$ correlations are relatively stronger in N(2, 0.2).
        \item Correlations among radii, $\Lambda$, $f$-mode frequencies and damping times are similar in both cases and they are similar to that of cold NSs \cite{Pradhan_Chatterjee_2022prc}.
    \end{itemize}
\end{enumerate}

\subsubsection{\texorpdfstring{$\chi$}{Lg}EFT + Astro + HIC}
The correlations on imposing $\chi$EFT + Astro + HIC constraints are shown for the two thermal cases in figs. \ref{fig:correlation_N(1, 0.4)_HIC} and \ref{fig:correlation_N(2, 0.2)_HIC}. We describe the key results and differences from $\chi$EFT + Astro filters below.
\begin{enumerate}
    \item \textbf{Correlations among nuclear saturation properties:} 
    \begin{itemize}
        \item Correlation between $J_{sym}$ and $L_{sym}$ gets weaker.
        \item Anti-correlation between $m^*/m$ and $n_{sat}$ becomes relatively stronger.
        \item $K_{sat}$ and $L_{sym}$ now have moderate correlations with $m^*/m$.
    \end{itemize}
    
    \item \textbf{Correlations of nuclear saturation properties with NS observables:} 
    \begin{itemize}
        \item $m^*/m$ anti-correlation with maximum mass remains strong for both cases but they are relatively weaker than those of correlations without HIC constraints.
        \item Masses, radii, $f$-mode frequencies and damping times of $\rm{1.4M_{\odot}}$ PNSs now have poor correlation with effective mass. %\sout{however its anti-correlation with $\rm{T_{c,1.4M_{\odot}}}$ for both the configurations get a little stronger and become moderate for both unlike that of constraints without HIC.}
        \item Correlations/anti-correlations of effective mass with properties ($R$, $\Lambda$, $\nu_f$, $\tau_f$) of canonical $\rm{2M_{\odot}}$ PNSs become moderate for both cases.
        \item Correlations/anti-correlations of $n_{sat}$ with observables ($R$, $\Lambda$, $\nu_f$, $\tau_f$) of canonical $\rm{1.4M_{\odot}}$ PNSs become stronger for both cases. However, its correlation with canonical $\rm{2M_{\odot}}$ hot NS observables remain weak.
        %\item $L_{sym}$ now shows moderate anti-correlation with $\rm{T_{c,1.4M_{\odot}}}$ for both the thermal cases.
        %\item Correlation of $K_{sat}$ with $\rm{T_{c,1.4M_{\odot}}}$ for both the cases remain unaffected.
    \end{itemize}

    \item \textbf{Correlations among NS observables:}
    \begin{itemize}
       \item Correlations/anti-correlations of $\rm{M_{max}}$ with observables of $\rm{1.4M_{\odot}}$  hot NSs become weaker for both cases.
      %\item \nb{Correlation of $\rm{T_{c,1.4M_{\odot}}}$ with $\rm{T_{c,2M_{\odot}}}$ for N(2, 0.2) remains stronger.}
      % \item Masses, radii, $f$-mode frequencies and damping times of $\rm{1.4M_{\odot}}$ configuration are now strongly correlated/anti-correlated with its central temperature.
    \end{itemize}
\end{enumerate}
%\newpage
\onecolumngrid

\begin{figure}[H]
    \centering
    \includegraphics[scale = .4]{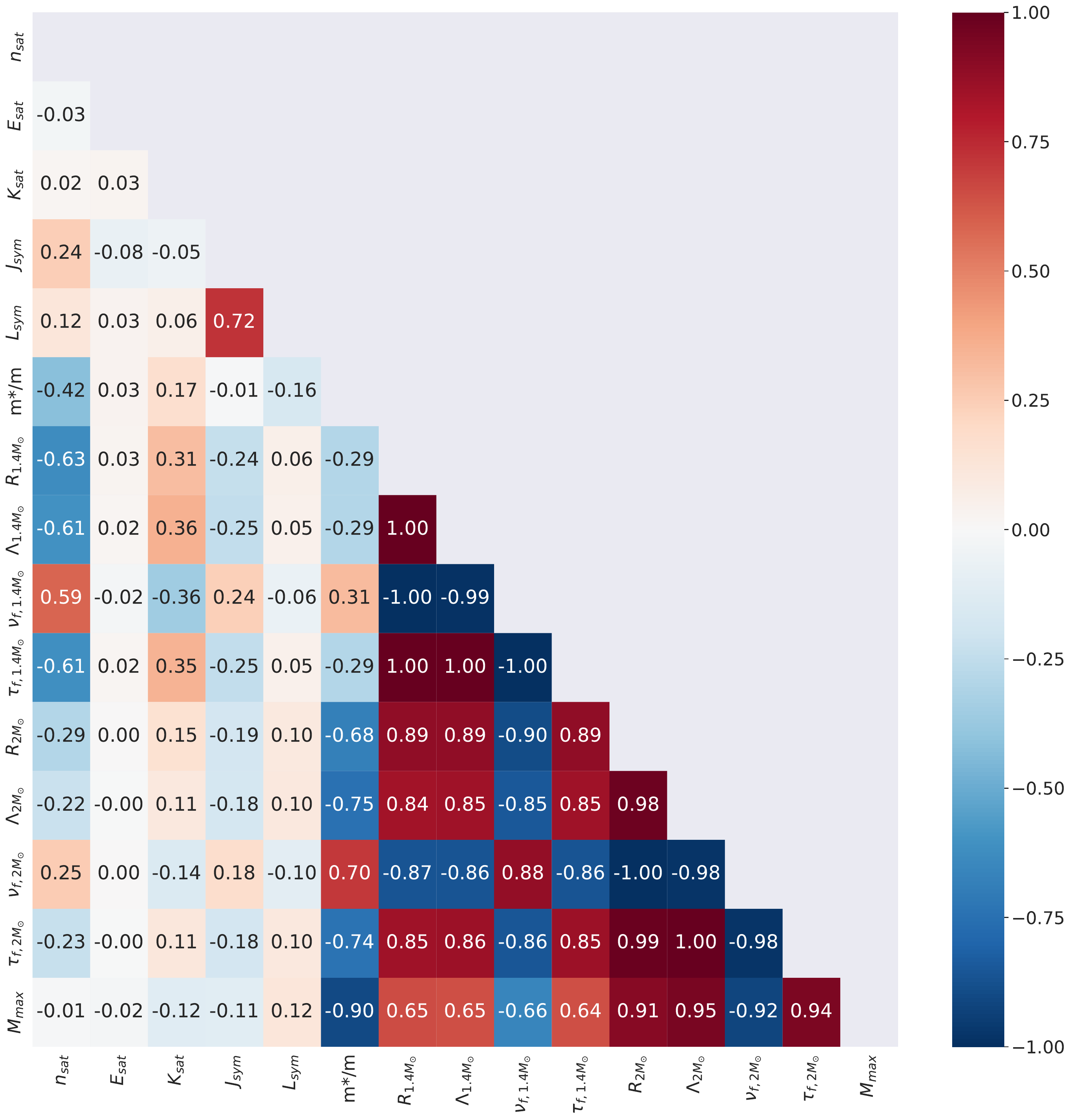}
    \caption{Correlations among saturation properties and observables for the state $N(1, 0.4)$ for $\chi$EFT + Astro constraints.}
    \label{fig:correlation_N(1, 0.4)}
\end{figure}

\begin{figure}[H]
    \centering
    \includegraphics[scale = .4]{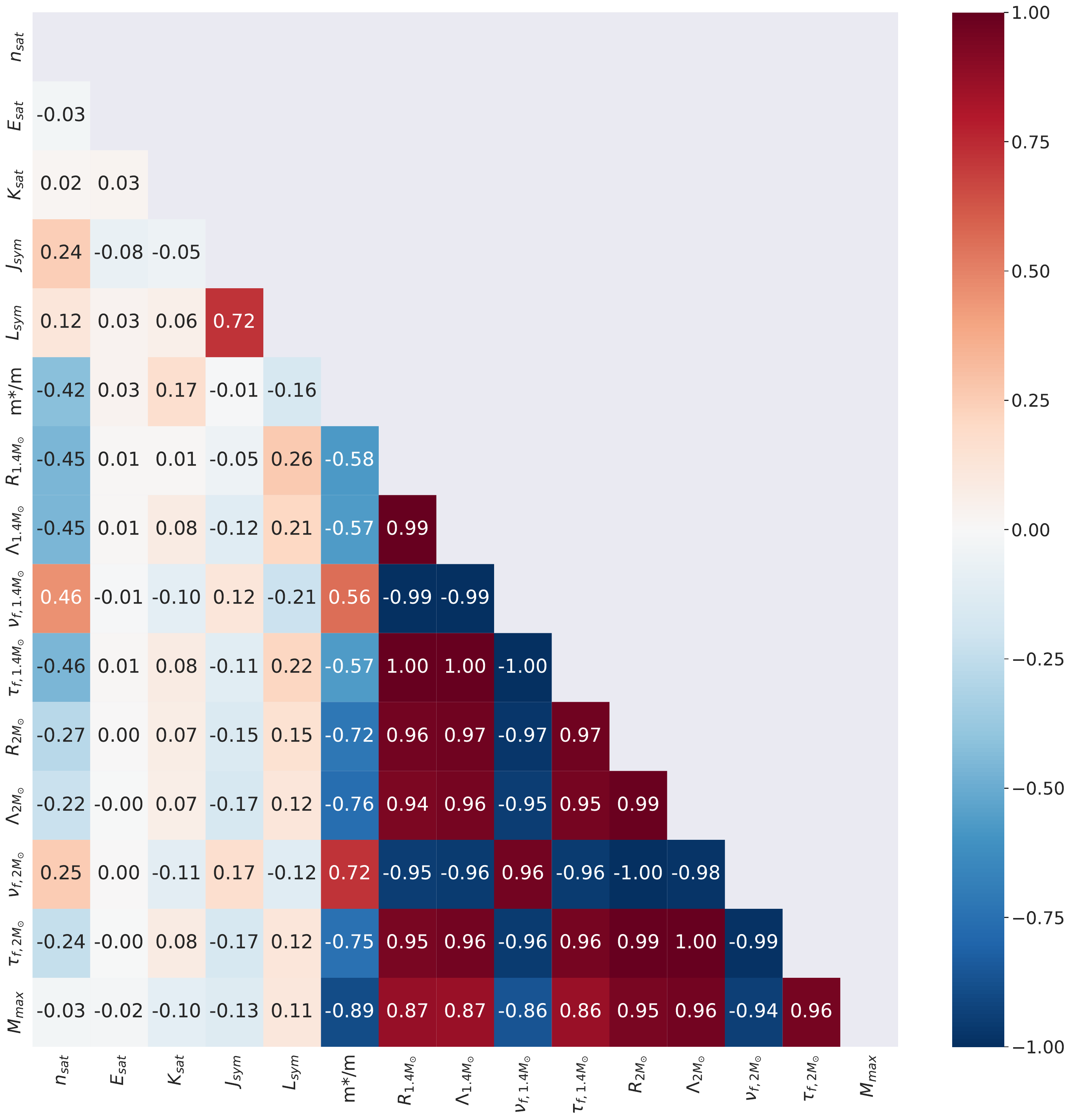}
    \caption{Correlations among saturation properties and observables for the state $N(2, 0.2)$ for $\chi$EFT + Astro constraints.}
    \label{fig:correlation_N(2, 0.2)}
\end{figure}

\begin{figure}[H]
    \centering
    \includegraphics[scale = .4]{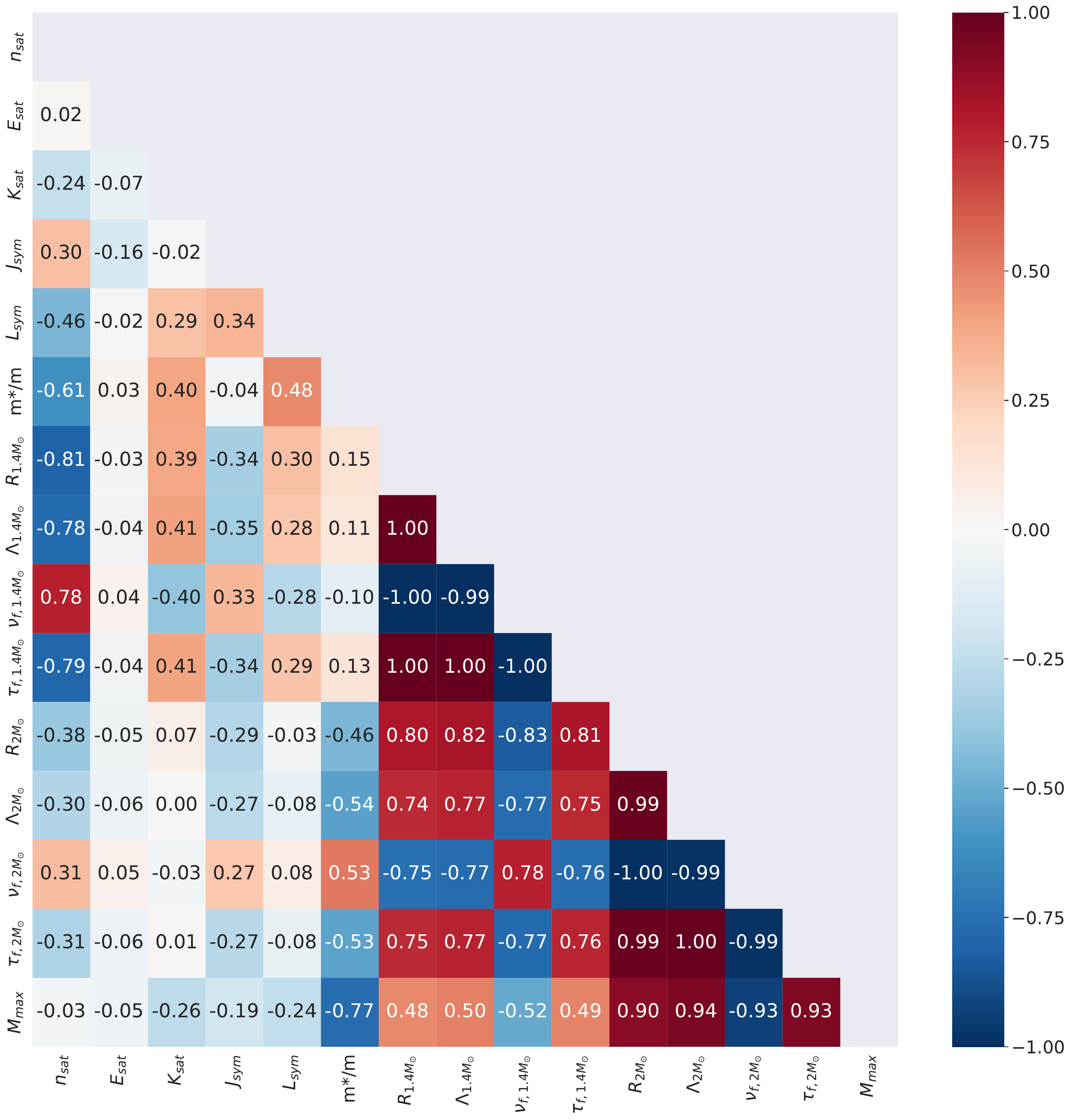}
    \caption{Correlations among saturation properties and observables for the state $N(1, 0.4)$ for $\chi$EFT + Astro + HIC constraints.}
    \label{fig:correlation_N(1, 0.4)_HIC}
\end{figure}

\begin{figure}[H]
    \centering
    \includegraphics[scale = .4]{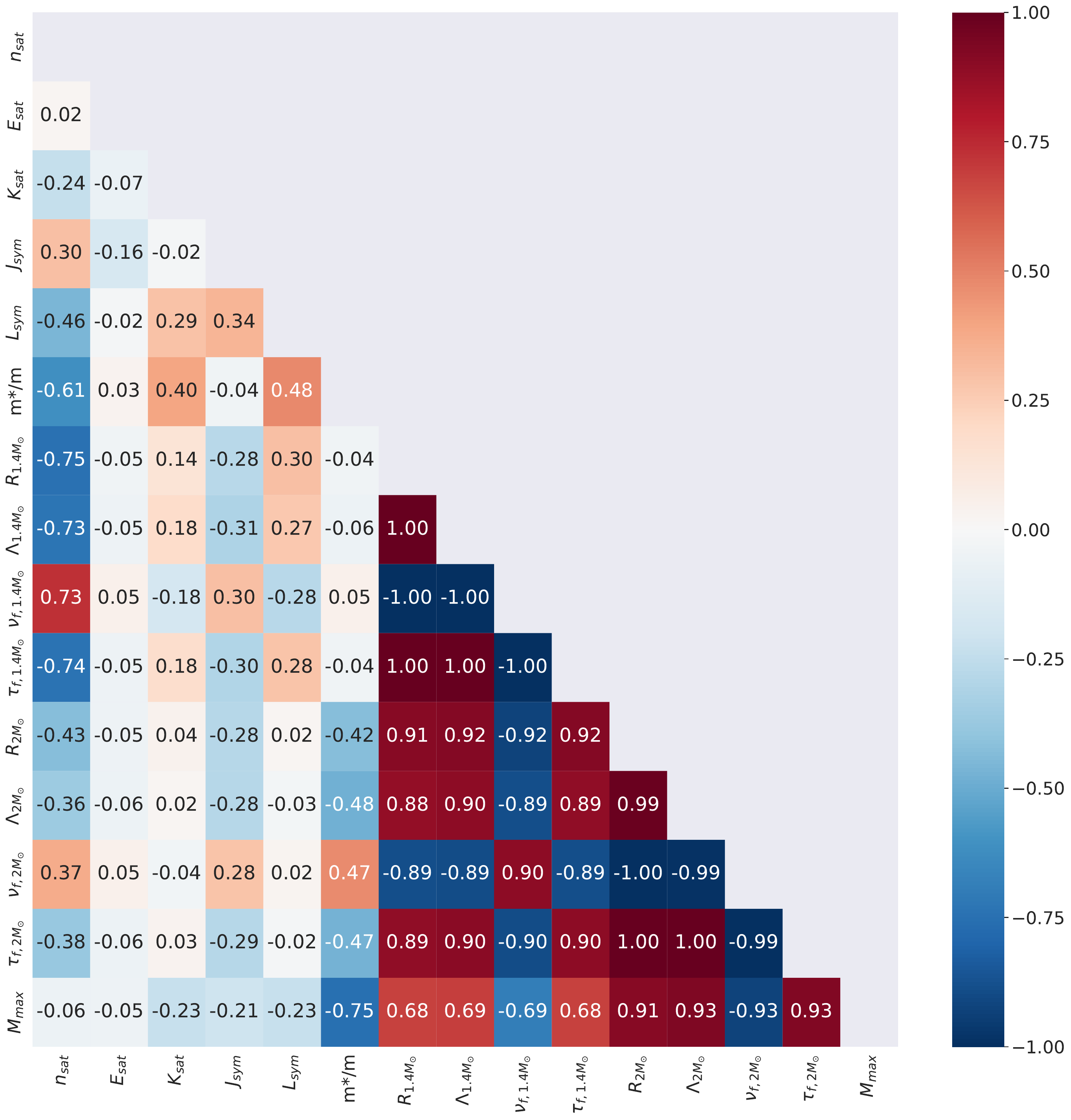}
    \caption{Correlations among saturation properties and observables for the state $N(2, 0.2)$ for $\chi$EFT + Astro + HIC constraints.}
    \label{fig:correlation_N(2, 0.2)_HIC}
\end{figure}

\newpage
\twocolumngrid
\section{Discussions}
\label{sec:discussions}

\textit{Summary of the results:} 
In this work, we develop a systematic and consistent formalism for nucleonic equations of state in neutron stars at finite temperature, by extending a cold matter EoS whose parameter space of allowed uncertainties of the model is restricted by demanding compatibility with current nuclear experimental data, multi-messenger astrophysical observations as well as heavy-ion physics. The developed framework was then applied to the calculation of polar $f$-modes and $p_1$-modes within a general relativistic scheme. The full ranges of allowed nuclear saturation properties for nucleonic matter used in our work allowed for the extensive study of correlations of the nuclear and thermal properties on macroscopic and oscillation properties of hot NSs.  We specifically chose a low entropy high charge fraction state and a high entropy deleptonized low charge fraction state and compared it with cold $\beta$ equilibrated NSs. 
\\

On including thermal contributions to the EoS, we found that increasing $S/A$ leads to higher core temperatures while $Y_Q$ has an opposite effect and deleptonized configurations are more compact. Increasing $S/A$ results in larger radii and flatter mass-radius relations. These observations are consistent with previous studies~\cite{Raduta_Oertel_2020mnras,Thapa_Beznogov_2023prd}. The $f$- and $p_1$-mode frequencies calculated in a full general relativistic scheme are found to be the highest for the cold configuration and decrease with an increase in entropy per baryon and charge fraction.
\\

On imposing $\chi$EFT, Astro and HIC constraints, we found that nuclear saturation parameters affect temperature profiles at intermediate and high densities. There are overlapping regions in the posteriors for $M-R$, $f$-mode frequency and damping timescales for high masses among $N(T=0),~N(1, 0.4),~N(2, 0.2)$ states. 
From the correlation study of hot NSs, we found on imposing $\chi$EFT and Astro constraints that the most important saturation properties having effect on astrophysical observables, such as maximum mass, radii, tidal deformation as well as $f$-mode characteristics, are found to be $m^*/m$ followed by $n_{sat}$.  
 However their correlations are qualitatively different depending on the thermal state after imposing $\chi$EFT + Astro constraints and $N(2, 0.2)$ state seems to resemble closely to that of cold NS results~\cite{Ghosh_Chatterjee_2022epja}. After applying HIC constraints, we found that correlations of $m^*$ with NS observables are significantly weaker. These findings are qualitatively similar to the results in cold $\beta$ equilibrated matter case~\cite{Ghosh_Chatterjee_2022epja,Pradhan_Chatterjee_2022prc}.
 %\nb{They do get weaker but the correlation numbers after imposing HIC are qualitatively similar to that of Suprovo's results for cold NSs, except it did not have $f$-modes. You can have a look at the fig. 12 in his paper} \dc{This is not true. Comparing Fig. 22 with 24, and 23 with 25 in your paper, it is evident that on imposing HIC the correlations of $m^*$ with observables get significantly weaker.}
 Hence, the interplay among saturation properties and astrophysical observables depends not only on the thermal configurations considered but also on the constraints imposed.
It should be noted that previous study of cold $\beta$ equilibrated NSs by Pradhan et al.~\cite{Pradhan_Chatterjee_2022prc} found strong correlations/anti-correlations of effective nucleon mass with all astrophysical observables but did not include constraints from $\chi$EFT and HIC. 
\\

Using the posteriors from the constraint filters, we further investigated the role of nuclear saturation parameters on the universal relations for hot NSs, in particular Love-compactness and $f$-Love relations. We concluded that, hot NS configurations at higher $\Lambda$ values show significant deviations in $C$-Love and $f$-Love relations compared to the cold $\beta$ equilibrated case, although these variations are more prominent in the former case. Further investigation confirmed that these URs are mostly insensitive to nuclear saturation properties and is mainly affected by charge fraction variation, which is a measure of out-of-$\beta$ equilibrium condition.
\\

\textit{Comparison with other works:} 
There exist other EoS models in the RMF scheme that have been extended to finite temperature and tested against current constraints. In the extension of the FSU2H parametrization based on the RMF scheme to finite temperature~\cite{Kochankovski_Ramos_2022mnras}, nucleonic and hyperonic matter have been considered. The resulting FSU2H* model is compatible with astrophysical constraints (maximum mass of 2 M$_{\odot}$ and radii below 13 km) as well as saturation properties of nuclear matter and finite nuclei, and pressure constraint from heavy-ion collisions. 
Alford et al.~\cite{Alford2023} also constructed QMC-RMF EoS models for finite temperature NSs, informed by $\chi$EFT and astrophysical constraints. These works demonstrate the strong influence of thermal corrections on the EoS and thermodynamic properties. However, they only consider fixed RMF parametrizations and therefore the role of the nuclear parameters in governing NS observable properties is not evident from such studies, but this is possible with a systematic variation of nuclear parameters as done in our work. Further, as these calculations are performed for isothermal and not isentropic conditions, it is not easy to estimate their effects on astrophysical observables.
\\

It is interesting to note that, several numerical simulations of CCSN, PNS as well as BNS mergers, employing different EoS models, have also reported the influence of the nucleon effective mass~\cite{Raithel_Paschalidis_2023prd,Fields_Prakash_2023apjl,Andersen_Zha_2021apj,Schneider_Constantinou_2019prc}. Using a selection of purely nucleonic EoS models from CompOSE (including Skyrme, covariant density functional models, CMF) in agreement with constraints from astrophysical observations and nuclear data, Raduta et al.~\cite{Raduta_Nacu_2021epja}, reviewed thermal properties relevant for PNS and BNS mergers and concluded that they are strongly influenced by the effective nucleon mass. This was further corroborated by using a covariant density functional theory with density dependent couplings employing the posteriors of a Bayesian inference (DDB*), which encompasses a large number of EoS models~\cite{Raduta_Beznogov_2024_plb}. 
These EoSs were also employed in
 Thapa et al.~\cite{Thapa_Beznogov_2023prd} to calculate the $f$- and $p_1$-mode frequencies within Cowling approximation. Although the thermal and composition effects on the oscillation modes were studied for the selection of EoSs from CompOSE, the correlations were calculated within the cold EoS approximation, and therefore did not provide any information about the thermal conditions.
\\

A few recent works by Ghosh et al.~\cite{Ghosh_Shaikh_2024} and Kumar et al.~\cite{Kumar_Thakur_2024mnras} studied oscillations within Cowling approximation including neutrino contributions for hot nucleonic and hybrid neutron stars. It should be noted that Burgio et al.~\cite{Burgio_Ferrari_2011prd} used a model where core and envelope entropy per baryon are different, and the former rises and then falls and the latter falls along with deleptonization during its evolution, to study $g_1$-, $f$- and $p_1$-modes using a BHF model extended to finite temperature. In our work, the entropy per baryon has been kept fixed throughout the star in our study. \\

In the recent work by Tsiopelas et al.~\cite{Tsiopelas_Sedrakian_2024}, which employed a density dependent covariant density functional model, the effect of $L_{sym}$ was found to be diminishing with increasing temperature. In our work using NL-RMF formalism we found that after applying $\chi$EFT + Astro filters, $L_{sym}$ has no impact on macroscopic or oscillation properties for $N(1, 0.4)$ configuration. On the other hand for $N(2, 0.2)$ configuration, these correlations are qualitatively similar to that of cold NS results~\cite{Ghosh_Chatterjee_2022epja}.
\\

Previous studies have indicated deviations of $\bar{I}-\bar{Q}$ relations in hot environment~\cite{Marques_Oertel_2017prc,Raduta_Oertel_2020mnras,Martinon_Maselli_2014prd,Lenka_Char_2019jpg}. A few studies~\cite{Raduta_Oertel_2020mnras,Khadkikar_Raduta_2021prc} showed that universality is restored if finite $T$ configurations for different EoS with same $S/A$ and electron fraction are considered. PNS simulations by Sotani et al.~\cite{Sotani_Kuroda_2017prd} explored the prospect of measurement PNS mass and radius using simultaneous measurements of $f$- (or $p_1$-) and $w_1$-mode frequencies by relating those to average mass density and stellar compactness respectively. Torres-Forn\'e et al.~\cite{Torres2019} have demonstrated how $g_1$ and $f$-mode frequencies scale with surface gravity in their PNS simulations. In our article, we have shown that $C$-Love and $f$-Love URs are mostly not sensitive to nuclear saturation properties but they change depending on the thermal state. In particular charge fraction significantly affects these deviations and they occur at high $\Lambda$-values (or low compactness). A recent simulation~\cite{Guedes_Lau_2024} has also shown that due to thermal effects deviations from URs i.e., $f$-$\sqrt{M/R^3}$, $\bar{I}-\Lambda-\bar{Q}$, $f$-Love occur during PNS evolution at earlier stages where the configuration has large tidal deformability. It should be noted that in the case of proto-neutron stars, in addition to thermal effects one must also take into account the effect of strong magnetic fields which are known to break the universality of $\bar{I}-$Love $-\bar{Q}$ relations~\cite{Haskell_Ciolfi_2013mnras}.
\\

\textit{Limitations and future work:}
It is worth mentioning that, we define the modes following the classification by Cowling~\cite{Cowling}. However, a recent investigation done in~\cite{Rodriguez2023}, considering the nucleonic EoS models, demonstrates that in the initial phase (within  0.4 s) of the PNS evolution, the Cowling classification scheme does not hold but after 0.4 sec the classification of oscillation modes with different approaches are in very good agreement. We further note that neutrinos have not been included in this work. Neutrinos may not remain free streaming for the entire temperature regime considered in this study, however we restrict ourselves to this approximation for this work and defer calculations including neutrino trapping to our forthcoming studies. 
\\

This study can be extended to include non-nucleonic degrees of freedom such as hyperons, delta baryons and quark matter in hot EoS that satisfy microscopic, experimental and astrophysical constraints. These new degrees of freedom are known to soften the EoSs and might also lead to decrease in core temperature due to entropy being contributed by these new species. \\

It would also be interesting to include the effect of neutrinos as they are expected to be in equilibrium with the thermal/isentropic bath at high temperatures.
Previous studies on BNS mergers~\cite{Alford_2018,Alford_2021} have shown that in the temperature regime $T \sim 1 - 5$ MeV, nuclear matter is transparent to neutrinos due to their long mean free path. At high temperatures (beyond $\sim 5$ MeV) neutrinos may be trapped, depending on the neutrino mean free path at a given density. As a first step, in this work we developed the formalism for the $\nu$-transparent case (valid for temperatures upto $5$ MeV) and plan to extend this to include trapped neutrinos in a future work.
\\

\acknowledgements
The authors are grateful to Suprovo Ghosh for providing the posterior data that was used in this work. D.C. would like to thank Adriana R. Raduta and Mikhail V. Beznogov for their warm hospitality at IFIN-HH, Bucharest which led to several interesting discussions related to this work. D.C. would also like to thank Prof. J. Schaffner-Bielich and his students at ITP Frankfurt for insightful discussions on finite temperature models of neutron stars. D.C. is grateful to Prof Mark Alford, Alexander Haber and their group at Washington University in St. Louis for their hospitality and fruitful exchanges. The authors also thank Prof Luciano Rezzolla for pointing them to his relevant works. N.B., B.K.P. and D.C. acknowledge the usage of the IUCAA HPC computing facility for numerical calculations of oscillation modes.

\section{Data availability}
The posterior data of both the nuclear saturation properties and the coupling constants obtained after imposing $\chi$EFT + Astro and $\chi$EFT + Astro + HIC constraints which support this work can be found at~\cite{coupling_constants}. Alternatively the posterior sets of only nuclear saturation properties are available in the arXiv ancillary files of previously published work of Ghosh et al.~\cite{Ghosh_Chatterjee_2022epja}.

\appendix

\section{Effect of entropy per baryon on \texorpdfstring{$f$}{Lg}-Love relations} 
\label{appendix:f-Love_entropy}

\begin{comment}
    \begin{figure}[H]
    \centering
    \includegraphics[width=1\linewidth]{Real_Mw_Lambda_s_variation.pdf}
    \caption{Effect of $S/A$ on $\rm{Real}$ ($M\omega_f)-\Lambda$ universality. Nuclear saturation properties are same as in fig. \ref{fig:T_vs_nb}. $S/A$ has been varied between 1-2 in steps of 0.01.}
    \label{fig:Real(Mw)_Lambda_s_variation}
\end{figure}

\begin{figure}[H]
    \centering
    \includegraphics[width=\linewidth]{Im_Mw_Lambda_s_variation.pdf}
    \caption{Effect of $S/A$ on $\rm{Im}$ ($M\omega_f)-\Lambda$ universality. Nuclear saturation properties are same as in fig. \ref{fig:T_vs_nb}. $S/A$ has been varied between $1-2$ in steps of 0.01.}
    \label{fig:Im(Mw)_Lambda_s_variation}
\end{figure}
\end{comment}

\begin{figure}[H]
    \centering
    \begin{subfigure}[b]{0.41\textwidth}
        \includegraphics[width=\linewidth]{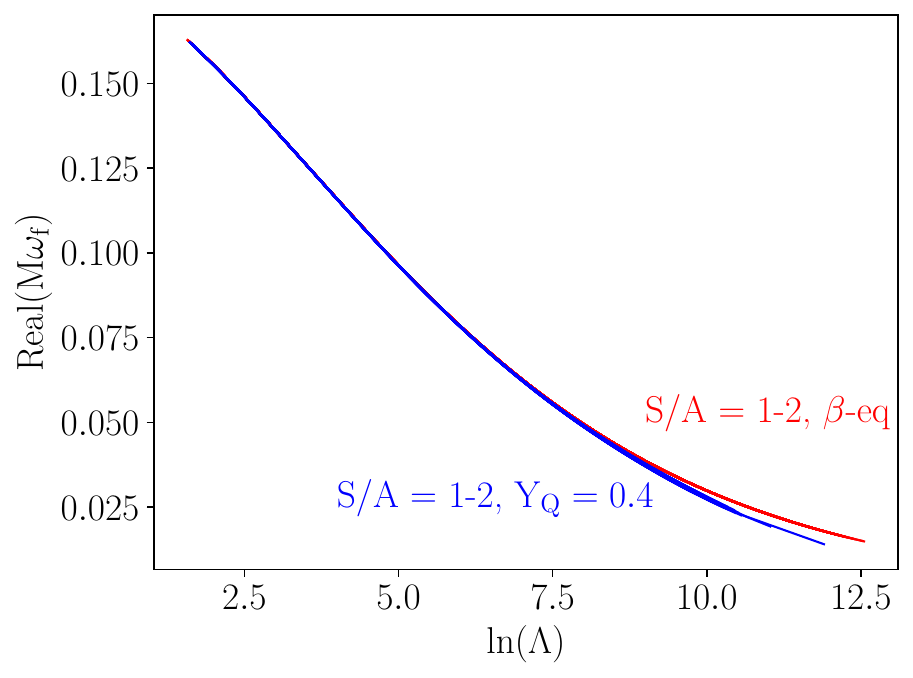}
        \caption{Real($M\omega)~v/s~\rm{ln(\Lambda)}$}
        \label{fig:Real(Mw)_Lambda_s_variation}
    \end{subfigure}
    \hfill
    \begin{subfigure}[b]{0.38\textwidth}
        \includegraphics[width=\linewidth]{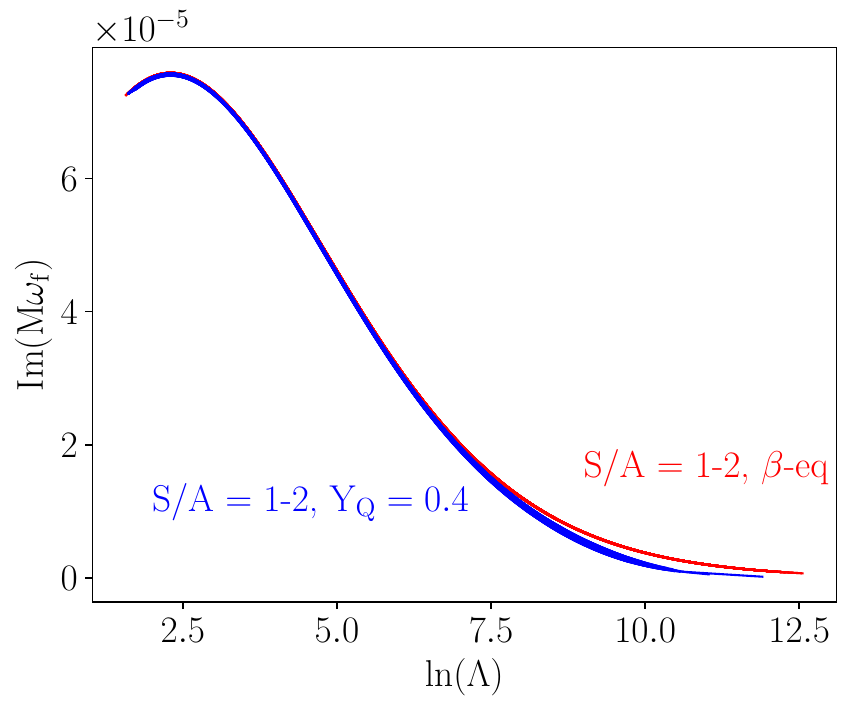}
        \caption{Im($M\omega)~v/s~\rm{ln(\Lambda)}$}
        \label{fig:Im(Mw)_Lambda_s_variation}
    \end{subfigure}
    \caption{Effect of entropy per baryon on $f$-Love universality.}
    \label{fig:Mw_Lambda_s_variation}
\end{figure}

We fix the nuclear saturation properties as in fig. \ref{fig:T_vs_nb}. Figs. \ref{fig:Real(Mw)_Lambda_s_variation} and \ref{fig:Im(Mw)_Lambda_s_variation} show effect of changing $S/A$ between $1-2$ in $Y_Q=0.4$ and $\beta$ equilibrated states.
From fig.~\ref{fig:Real(Mw)_Lambda_s_variation}, we see that $Y_Q=0.4$ case shows deviations from $\beta$ equilibtrated hot NSs for higher $\Lambda$ and they are overlapping otherwise.  This is in contrast to $\Lambda-C$ URs shown in fig. \ref{fig:Lambda_C_s_variation} where we did not get any overlapping region. The relation between $\rm{Im}$($M\omega)$ and $\ln(\Lambda)$ are shown in fig. \ref{fig:Im(Mw)_Lambda_s_variation} also establishes deviations at high $\Lambda$ values. There are also qualitatively negligible deviations for both real and imaginary parts across the entropy per baryon range considered as compared to $\Lambda-C$ URs.

These deviations shown between states with $\beta$ equilibration and $Y_Q=0.4$ clearly shows that at higher $\Lambda$ universality is broken by out-of-$\beta$ equilibrium effects. 
%We have discussed effect of $Y_Q$ on $f$-Love universality in Appendix~\ref{appendix:f-Love}. 
The spreads however in figs. \ref{fig:Real(Mw)_Lambda_s_variation} and \ref{fig:Im(Mw)_Lambda_s_variation} are qualitatively much lower than that of $\Lambda-C$ relations shown in fig. \ref{fig:Lambda_C_y_variation} where deviations are observed across the entire range of dimensionless tidal deformability. \\

\section{Effect of charge fraction on \texorpdfstring{$f$}{Lg}-Love relations} \label{appendix:f-Love_yq}
\begin{figure}[H]
    \centering
    \begin{subfigure}[b]{0.4\textwidth}
        \includegraphics[width=\linewidth]{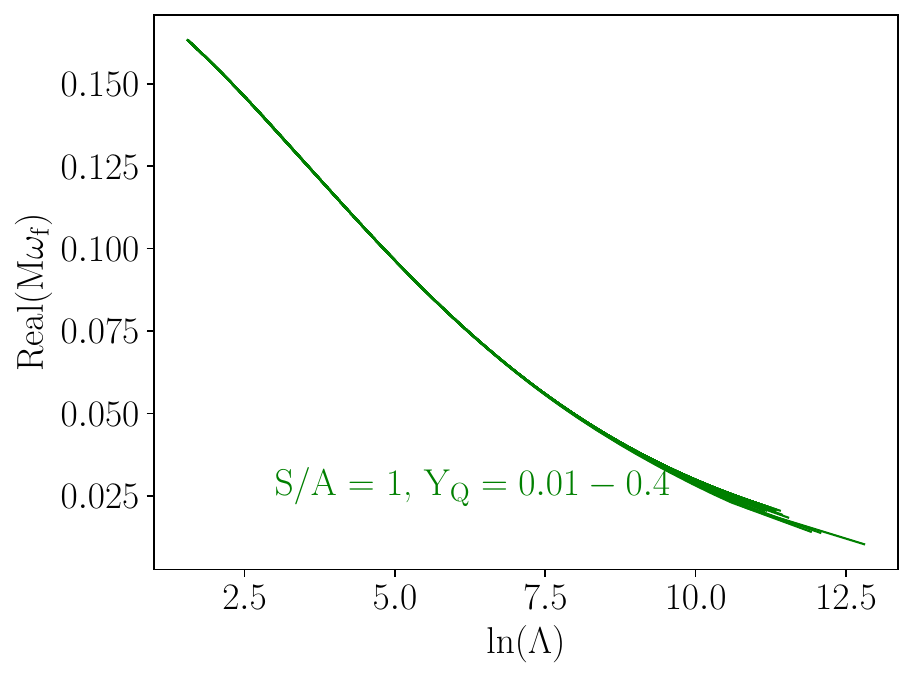}
        \caption{Real($M\omega)~v/s~\rm{ln(\Lambda)}$}
        \label{fig:Real(Mw)_Lambda_y_variation}
    \end{subfigure}
    \hfill
    \begin{subfigure}[b]{0.38\textwidth}
        \includegraphics[width=\linewidth]{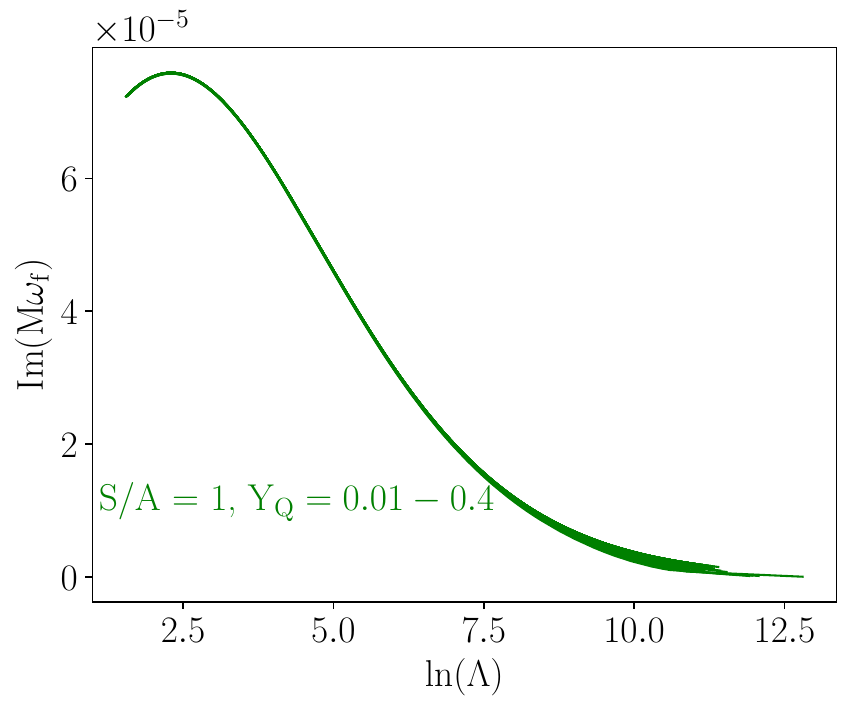}
        \caption{Im($M\omega)~v/s~\rm{ln(\Lambda)}$}
        \label{fig:Im(Mw)_Lambda_y_variation}
    \end{subfigure}
    \caption{Effect of charge fraction on $f$-Love universality.}
    \label{fig:Mw_Lambda_y_variation}
\end{figure}

Here we fix entropy per baryon at $S/A=1$ and vary $Y_Q$ from 0.01 - 0.40. We see in figs.~\ref{fig:Real(Mw)_Lambda_y_variation} and~\ref{fig:Im(Mw)_Lambda_y_variation} that this has no effect on both real and imaginary parts of $f$-Love relations for low $\Lambda$ values and at higher values there are small spreads. These deviations appear for $\Lambda>e^8\sim3000$. The conclusion is that charge fraction is responsible for deviations in the universality like $C$-Love relations albeit to a much lesser degree and only at higher $\Lambda$.

\bibliography{barman.bib}% Produces the bibliography via BibTeX.

\end{document}